\PassOptionsToPackage{pdfpagelabels=false}{hyperref}

\documentclass[fleqn,usenatbib,useAMS]{mnras}

\usepackage{graphicx}
\usepackage{amsmath}
\usepackage[dvipsnames]{xcolor}  
\usepackage{amssymb}
\usepackage{color}
\usepackage{soul}
\usepackage[figure,figure*,table,table*]{hypcap}

\input{macros.sty}

\title[Maturing Satellite Kinematics]
{Maturing Satellite Kinematics into a Competitive Probe of the Galaxy-Halo Connection}
\author[J.~U.~Lange et al.]
{Johannes~U.~Lange$^{1, 2}$\thanks{email: johannesulf.lange@yale.edu}, Frank~C.~van~den~Bosch$^1$, Andrew~R.~Zentner$^3$\newauthor Kuan~Wang$^3$ and Antonio~S.~Villarreal$^3$\\
	$^1$Department of Astronomy, Yale University, P.O. Box 208101, New Haven, CT 06511, USA\\
	$^2$Kavli Institute for Theoretical Physics, University of California, Santa Barbara, CA 93106, USA\\
	$^3$Department of Physics and Astronomy \& Pittsburgh Particle Physics, Astrophysics, and Cosmology Center (PITT PACC),\\University of Pittsburgh, Pittsburgh, PA 15260, USA}

\pubyear{2018}

\begin{document}
	
	\date{Accepted xxx. Received xxx}
	
	\label{firstpage}
	\pagerange{\pageref{firstpage}--\pageref{lastpage}}
	
	\maketitle
	
	\begin{abstract}
		The kinematics of satellite galaxies moving in a dark matter halo are a direct probe of the underlying gravitational potential. Thus, the phase-space distributions of satellites represent a powerful tool to determine the galaxy--halo connection from observations. By stacking the signal of a large number of satellite galaxies this potential can be unlocked even for haloes hosting a few satellites on average. In this work, we test the impact of various modelling assumptions on constraints derived from analysing satellite phase-space distributions in the non-linear, 1--halo regime. We discuss their potential to explain the discrepancy between average halo masses derived from satellite kinematics and gravitational lensing previously reported. Furthermore, we develop an updated, more robust analysis to extract constraints on the galaxy--halo relation from satellite properties in spectroscopic galaxy surveys such as the SDSS. We test the accuracy of this approach using a large number of realistic mock catalogues. Furthermore, we find that constraints derived from such an analysis are complementary and competitive with respect to the commonly used galaxy clustering and galaxy--galaxy lensing observables.
	\end{abstract}
	
	
	\section{Introduction}
	
	In 1933, Fritz Zwicky was one of the first astronomers to find evidence for the existence of dark matter. He based his conclusions on the observed velocity dispersion of galaxies in the Coma cluster. Because galaxy kinematics directly probe the underlying gravitational potential, \cite{Zwicky_33} concluded that most of the matter inside Coma could be dark. 
	
	Today, the $\Lambda$ cold dark matter ($\Lambda$CDM) paradigm is our current best cosmological model for the formation and evolution of cosmic structure. In this scenario, dark matter haloes provide the potential wells that enable the formation of galaxies. Hence, constraining the link between galaxies and dark matter haloes, which goes by the catch-all name `halo-occupation modelling', can provide valuable insight regarding galaxy formation. Our current best constraints on this galaxy-dark matter connection come from galaxy clustering \citep[e.g.,][]{vdBosch_07, Zehavi_11, Hearin_13c, Guo_15a, Guo_15b, Guo_16, Zentner_16, Moster_18, Campbell_18}, galaxy-galaxy lensing \citep[e.g.,][]{Sheldon_09a, Sheldon_09b, Zu_15, Zu_16, Mandelbaum_16, Sonnenfeld_18a}, and combinations thereof \citep[e.g.,][]{Cacciato_09, Leauthaud_12, Cacciato_13, Wibking_17, DES_17}. Another popular method used in halo occupation modelling is abundance matching, which postulates a tight, monotonic relation between halo mass and galaxy luminosity or stellar mass. Abundance matching can be applied to individual galaxies, in which case one talks of subhalo abundance matching \citep[e.g.,][]{Vale_04, Vale_06, Guo_10, Hearin_13c}, or to galaxy groups \citep[e.g.,][]{Yang_05, Yang_07, Yang_08, Yang_09, Weinmann_06}, in which case one only uses host haloes. See \citet{Wechsler_18} for an excellent review of the galaxy-halo connection.
	
	Somewhat surprisingly, despite being the original method used by Zwicky to reveal the presence of dark matter, satellite kinematics has hitherto been little utilized in halo occupation modelling. Although various studies have used the kinematics of satellite galaxies as tracers of their dark matter potential wells \citep{McKay_02, Prada_03, Brainerd_03, Conroy_07, Norberg_08, More_09a, More_09b, More_11, Li_12, Wojtak_13}, very few have done so within the framework of halo occupation modelling. One reason for this is that most previous studies applied highly restrictive isolation criteria to select central (or `host') and satellite galaxies and, consequently, these samples can no longer be considered representative of the galaxy population as a whole. In addition, the strictness severely limited the total number of central-satellite pairs, and hence the signal-to-noise and dynamic range of the data. The goal of this paper is to rectify this situation and to promote satellite kinematics as a halo-occupation modelling tool on par with clustering and lensing.
	
	Following pioneering efforts by \cite{Erickson_87} and \cite{Zaritsky_93, Zaritsky_97}, the first attempt to use satellite kinematics to infer the galaxy--halo connection from a large redshift survey was by \cite{McKay_02}, who used the \textit{Sloan Digital Sky Survey} (SDSS) to study the velocity distribution of $1225$ satellites around $618$ central galaxy candidates. The main finding was that the average host halo masses scale roughly linearly with the luminosity of the central galaxies, confirming previous results from \cite{McKay_01} using galaxy-galaxy lensing. A similar study was performed by \cite{Prada_03} utilizing up to $2734$ satellites in the SDSS. They found that the velocity dispersion of satellites decreases with the distance to the central candidate, in good agreement with predictions from dark matter simulations. They also confirmed the results of \cite{McKay_02} that dark matter halo mass increases with the luminosity of the central. \cite{Brainerd_03} analysed the kinematics of $1556$ satellites in the Two-Degree Field Galaxy Redshift Survey (2dFGRS). They mostly confirmed previous results by \cite{McKay_02} and additionally found that the dark matter halo mass seems to be independent of luminosity for spiral galaxies. \cite{Conroy_07} expanded the use of satellite kinematics to higher redshifts, $z \sim 1$, using $\sim 1000$ satellites in the DEEP2 Galaxy Redshift Survey \citep[][]{Davis_03}. By comparing results from both DEEP2 and SDSS at lower redshifts they found the stellar-to-halo mass ratio to be mostly independent of stellar mass and redshift. \cite{Norberg_08} analysed the motions of $1003$ satellites around $571$ central candidates in 2dFGRS. By comparing their findings to the previous studies, they pointed out that most satellite kinematics studies thus far had produced quantitatively very discrepant results. They argued that these discrepancies arise from the use of different isolation criteria, and different estimators of the satellite velocity dispersion, and concluded that the interpretation of satellite kinematics data remains ``questionable'' unless one applies the same methods to a set of realistic mock galaxy catalogues. Finally, \cite{Wojtak_13} were the first to try and constrain the anisotropy of satellite orbits; by modelling both the line-of-sight velocities and projected distances of more than $10,000$ satellite galaxies in the SDSS Data Release 7 \citep[DR7][]{Abazajian_09}, with respect to their centrals, they were able to simultaneously constrain the halo masses and the orbital anisotropy of the satellites (but see below).
	
	A commonality among the satellite kinematic studies listed above is that they have been extremely conservative in their selection of centrals and satellites. In particular, centrals were selected to be significantly brighter than any other galaxy within their neighborhood (typically defined as a cylindrical region in redshift-space, see below), and satellites typically had to be at least 4 to 8 times fainter than their corresponding central. This not only severely limits the number of central-satellite pairs, it also introduces a selection bias against haloes with bright satellites. This in turn could introduce a significant bias in halo mass. These shortcomings were circumvented by \cite{vdBosch_04}, who introduced an iterative, adaptive method to select centrals and satellites. This drastically increased the number of central-satellite pairs, while simultaneously decreasing the fraction of interlopers (galaxies unassociated with the dark matter halo of the central) and increasing the dynamic range of the galaxy-halo connection probed.
	
	Another important point made by \cite{vdBosch_04} is the impact of mass-mixing: because there is no one-to-one relation between central luminosity or stellar mass and dark matter halo mass, stacking implies that one combines the kinematics of satellites orbiting in haloes of different masses. If this is not properly accounted for in the analysis, one can make large errors in the inferred (mean) masses, or, as in the case of \cite{Wojtak_13}, the inferred orbital anisotropy \citep[see also][]{Becker_07, Norberg_08}. \cite{More_09a} developed a method that accounts for this mass mixing when stacking central galaxies. In particular, they demonstrated that one can actually measure the amount of scatter in halo mass by comparing the satellite kinematics obtained using different weighting schemes.
	
	One of the most comprehensive studies to date constraining the global galaxy-halo connection from satellite kinematics is that by \cite{More_11}. This study is a continuation of the work by \cite{vdBosch_04} and \cite{More_09b, More_09a}. Using data of $\sim 6000$ satellite galaxies selected from a volume-limited subsample of the SDSS DR4 \citep[][]{Adelman-McCarthy_06}, covering the redshift range $0.02 \leq z \leq 0.071$, \cite{More_11} obtained constraints on the galaxy-dark matter connection that were qualitatively in good agreement with other investigations based on  clustering and/or lensing. Furthermore, \cite{More_11} confirmed previous findings by \cite{Conroy_07} that red centrals live in more massive haloes than blue centrals of the same luminosity \citep[also see][for splits by morphological type]{Norberg_08}. They also found the same trend using stellar mass samples, a result that has since been verified by other studies using gravitational lensing \citep{Velander_14, Mandelbaum_16, Zu_16}. However, despite this success, on a more quantitative level, the results of \cite{More_11} imply average halo masses that are significantly different (by a factor of two to three) than those inferred from clustering and/or galaxy-galaxy lensing \citep[see e.g.,][]{Dutton_10, Leauthaud_12, Mandelbaum_16}. Such a level of disagreement is well beyond what is considered acceptable for numerous applications of the galaxy-halo connection \citep[see e.g.,][]{Wechsler_18}, and is in dire need of an explanation.
	
	This is the first paper in a series in which we seek to significantly improve upon previous studies of satellite kinematics, and to develop a methodology that is both complementary to, and competitive with, other probes of the galaxy-dark matter connection, such as clustering and galaxy-galaxy lensing. In this first paper we identify potential shortcomings in the analysis by  \cite{More_09b, More_09a, More_11}, which we take as the starting point of our investigation. We use detailed mock galaxy redshift surveys to compute a proper covariance matrix, to be used in the analysis, and to demonstrate the importance of accounting for redshift incompleteness, in particular fibre-collisions, something that was not accounted for in any previous satellite kinematic study with the exception of \cite{Wojtak_13}. We also improve upon the treatment of interlopers, and use forward modelling to quantify small biases in the analytical model. Using realistic mock catalogues, we demonstrate that these biases can be properly accounted for, resulting in a method for analyzing satellite kinematics that yields unbiased constraints on the galaxy-dark matter connection that are both competitive with and complimentary to studies based on clustering or lensing. In Paper II (Lange et al., in prep.), we apply this improved method to the SDSS DR7, finding constraints on the galaxy-halo connection that are in good agreement with other independent studies, abating the tension with the analysis by \cite{More_11} eluded to above. 
	
	This paper is organized as follows. In \S\ref{sec:methodology} we start with a broad outline of the methodology. \S\ref{sec:galaxy-halo_connection} describes our model for how galaxies occupy dark matter haloes, while \S\ref{sec:mocks} describes our algorithm for creating realistic mock SDSS-like catalogues. The observables that we use to constrain the galaxy--halo connection are detailed in \S\ref{sec:observations} and a simplified analytical model for those observables, for a given galaxy--halo connection, is described in \S\ref{sec:model}. \S\ref{sec:analysis} describes the full analysis procedure, which we apply to mock catalogues with known input parameters in \S\ref{sec:application_mocks}. Finally, in \S\ref{sec:conclusion} we summarize our findings.
	
	Throughout this work we assume a flat $\Lambda$CDM cosmology with $\Omega_\rmm = 0.307$, $\Omega_\rmb = 0.048$, $\sigma_8 = 0.823$, and $h = 0.678$. Here, $h = H_0 / (100 \ \mathrm{km/s/Mpc})$ and $H_0$ is the Hubble constant \citep{Planck_14}. All magnitudes are given in the AB magnitude system. When appropriate, quantities like comoving distance, halo mass or luminosity are scaled by $h$ to make results independent of $H_0$. Finally, throughout this paper we use $r$ to denote 3D radii, and $R$ for projected 2D radii.
	
	\section{Methodology}
	\label{sec:methodology}
	
	This section provides a rough outline of our methodology. We first describe how to measure satellite kinematics from a galaxy redshift survey, and then detail how to use such data to constrain the galaxy-dark matter connection. A more detailed description of our methodology is given in \S\S\ref{sec:galaxy-halo_connection}--\ref{sec:analysis} below.
	
	The first step in measuring satellite kinematics from a galaxy redshift survey is the selection of centrals and satellites. The standard method of selecting centrals is based on a cylindrical isolation criterion; i.e., a galaxy is identified as a central if it is the brightest galaxy in some specified, cylindrical volume whose symmetry axis falls along the line-of-sight. Satellite galaxies are defined as those galaxies that fall within a similar cylindrical volume, centred on the central galaxy, that meet a certain number of criteria. Ideally this selection of centrals and satellites is optimized to maximize completeness and minimize impurities (see \S\ref{subsec:cylinders} for details).
	
	Using the centrals and satellites thus identified, the next step is to quantify the kinematics of satellite galaxies within the host haloes of their associated centrals. Since the typical number of satellite galaxies per central is small, except in nearby clusters, this requires stacking whereby one co-adds all central-satellite pairs for centrals specified by $n$ galaxy properties ${\calG_1,\calG_2,...,\calG_n}$ (e.g., luminosity, color, size). The satellite kinematics are then specified by the line-of-sight velocity distribution (LOSVD) $P(\dv |\calG_1, \calG_2,...,\calG_n)$, where 
	\begin{equation}\label{eq:dv}
		\dv = c \frac{(z_{\rm sat} - z_{\rm cen})}{1+z_{\rm cen}}\,,
	\end{equation}
	with $z_{\rm sat}$ and $z_{\rm cen}$ the observed redshifts of the satellite and central, and $c$ the speed of light. The summary statistic that is most often used in the study of satellite kinematics is the satellite velocity dispersion, $\sigma_{\rm sat}(\calG_1, \calG_2,...,\calG_n)$,  which characterizes the second moment of this LOSVD.
	
	The main goal of the analysis of satellite kinematics envisioned in this paper, is to constrain the galaxy-dark matter connection, which can be quantified in terms of the conditional probability $P(\calH | \calG_1, ..., \calG_n)$. Here $\calH$ is the halo property of interest, which typically will be some measure of halo mass. The aim of such an analysis is a translation of $\sigma_{\rm sat}(\calG_1, \calG_2,...,\calG_n)$,  or more generally $P(\dv |\calG_1, \calG_2,...,\calG_n)$, into corresponding constraints on $P(\calH | \calG_1, \calG_2,...,\calG_n)$. Ideally this is done using forward modelling based on mock galaxy redshift surveys. This process consists of the following steps:
	\begin{enumerate}
		\item Using Monte Carlo techniques, populate the dark matter halos in a dark-matter-only numerical simulation of a cosmological volume according to a particular realization of the model for the galaxy-halo connection, $P(\calH | \calG_1, \calG_2,...,\calG_n)$. 
		\item Place a virtual observer in the cosmological volume, define an angular coordinate system with respect to that observer, and compute for each mock galaxy the corresponding right ascension and declination, as well as its redshift (including peculiar velocity) and apparent magnitude.
		\item Apply the survey mask and sample selection of the redshift survey, and mimic, insofar as is practical, survey incompleteness effects such as fibre 
		collisions. 
		\item Apply the same central/satellite selection criteria as for the real data, and compute the corresponding $P(\dv |\calG_1, \calG_2,...,\calG_n)$ and/or  $\sigma_{\rm sat}(\calG_1, \calG_2,...,\calG_n)$.
		\item Compare to the data, and embed in a Bayesian inference engine to constrain the posterior distributions of the parameters that quantify $P(\calH | \calG_1, \calG_2,...,\calG_n)$. 
	\end{enumerate}

	This forward modelling approach has a number of advantages. First of all, with forward modelling it is straightforward to account for biases that may arise from details related to the selection of centrals and satellites, and for a variety of observational complications, such as fibre collisions. Secondly, it trivially allows one to use the full information from the LOSVD, $P(\dv |\calG_1, \calG_2,...,\calG_n)$, rather than just its second moment. 
	
	Unfortunately, a full forward modelling approach as outlined above is currently computationally infeasible. Tests show that we need $\sim 10^8$ model evaluations in order to reliably constrain the posteriors of our model parameters that quantify the galaxy-dark matter connection (see \S\ref{sec:galaxy-halo_connection}). Furthermore, each model evaluation requires of order 100 mocks to suppress the realization noise from a single SDSS-size mock. The construction of a single mock, and the subsequent computation of $P(\dv |\calG_1, \calG_2,...,\calG_n)$, is orders of magnitude too slow to allow for the construction of $\calQ(10^{10})$ mocks, and we therefore have to rely on a less computationally-intensive approximation to the aforementioned process.
	
	As in previous studies \citep[e.g][]{vdBosch_04, More_09b, More_11, Wojtak_13}, we therefore use an analytical model which relies on the Jeans equations to predict the satellite kinematics as a function of halo mass. However, we improve upon these previous studies in a number of ways: (i) we use a proper covariance matrix for all the data,  (ii) we account for fibre collisions and other survey selection effects, and (iii) we use a new method to correct for interlopers that does not rely on some ad-hoc assumption regarding the functional form of $P(\dv |\calG_1, \calG_2,...,\calG_n)$. We also use detailed forward modelling as described above to gauge the reliability of the analytical model. In particular, we characterize its systematic bias, and use that to correct the model. Finally, once we have obtained the posterior distribution for the model, we construct a set of mocks for the best-fit model, and use forward modelling to verify that no systematic error has seeped into our approach. If necessary, we modify our treatment for the bias and iterate this entire procedure until convergence. This strategy is an intermediate step, necessitated by computational limitations, between previous analyses and more complete forward modelling.
	
	Our model based on the Jeans equations only predicts the second moment of the full LOSVD. Comparison with the data therefore requires that we compute $\sigma_{\rm sat}(\calG_1, \calG_2,...,\calG_n)$ from the observed $P(\dv |\calG_1, \calG_2,...,\calG_n)$, which, as we discuss in \S\ref{subsec:accuracy_analytic}, is hampered by a systematic bias. Although our method is codified to correct for this bias, an obvious downside is that our analytical model does not use all information that is available from the detailed {\it shape} of the LOSVD. However, we can access some of that information by using a clever method, first proposed by \cite{More_09a}: by using two different weights for the central-satellite pairs (host-weighting and satellite-weighting, see \S\ref{subsec:observables}), we obtain two different estimates of the second moment, $\sigma_{\rm sat}(\calG_1, \calG_2,...,\calG_n)$. Their difference can be used to constrain higher-order moments of $P(\calH | \calG_1, \calG_2,...,\calG_n)$, such as the amount of scatter in the relation between halo mass and galaxy luminosity.
	
	\section{Galaxy-Halo Connection}
	\label{sec:galaxy-halo_connection}
	
	When modelling satellite kinematics, and when constructing mock galaxy catalogues from dark-matter-only simulations, one needs to characterize the abundance, demographics, and phase-space distributions of central and satellite galaxies as a function of halo mass. In this section we describe the model that we use to parametrize this `galaxy-halo connection'.
	
	\subsection{Galaxy Occupation Statistics}
	\label{sec:CLF}
	
	We model the galaxy occupation statistics using the conditional luminosity function \citep[CLF;][]{Yang_03}, $\Phi(L|M) \rmd L$, which describes the average number of galaxies with luminosities in the range $L \pm \rmd L/2$ that reside in a dark matter halo of virial mass $M$. The CLF is composed of a central and a satellite part,
	\begin{equation}
		\Phi_{\rm tot}(L | M) = \Phi_\rmc (L | M) + \Phi_\rms (L | M),
	\end{equation}
	where subscripts `c' and `s' refer to central and satellite, respectively. In what follows we describe each of these two components in turn.
	
	\subsubsection{Centrals}
	
	Each halo hosts at most one central galaxy, which can be either red or blue. We assume that the colour distribution is governed by halo mass \citep[see e.g.][]{More_11, Mandelbaum_16, Zu_16}, such that
	\begin{equation}
		\Phi_\rmc (L | M) = f_\rmr (M) \Phi_{\rmc, \rmr}(L | M)+ f_\rmb (M) \Phi_{\rmc, \rmb}(L | M)\,,
	\end{equation}
	where subscripts `r' and `b' refer to red and blue, respectively. Here $f_\rmr (M) = 1 - f_\rmb(M)$ is the probability that a central galaxy in a halo of mass $M$ is red. Throughout we assume that $f_\rmr$ scales linearly with halo mass according to
	\begin{equation}
		f_\rmr (M) = \min\left[1, \max(0, f_0 + \alpha_f \log M_{12})\right]\,,
	\end{equation}
	where $M_{12} = M / (10^{12} \Msunh)$. The CLFs of red and blue centrals are parametrized separately by independent log-normal distributions,
	\begin{equation}
		\Phi_{\rm c,i} (L | M) dL = \frac{\log e}{\sqrt{2\pi \sigma_{\rm c,i}^2}} \exp \left( -\frac{[\log L - \log \tilde L_{\rm c,i} (M)]^2}{2 \sigma_{\rm c,i}^2} \right) \frac{\rmd L}{L}.
	\end{equation}
	The subscript i=r,b serves to distinguish the parameters for red and blue galaxies. 
	The median luminosity $\tilde L_\rmc (M)$, is in turn parametrized by a broken
	power-law:
	\begin{equation}
		\tilde L_{\rm c,i} (M) = L_{0,{\rm i}} \frac{(M / M_{1,{\rm i}})^{\gamma_{1,{\rm i}}}}{(1 + M / M_{1,{\rm i}})^{\gamma_{1,{\rm i}} - \gamma_{2,{\rm i}}}}.
	\end{equation}
	Altogether, we have $12$ free parameters to describe how central galaxies populate dark matter haloes; $f_0$, $\alpha_f$, $L_{0, \rmr}$, $M_{1, \rmr}$, $\gamma_{1, \rmr}$, $\gamma_{2, \rmr}$, $\sigma_\rmr$, $L_{0, \rmb}$, $M_{1, \rmb}$, $\gamma_{1, \rmb}$, $\gamma_{2, \rmb}$, and $\sigma_\rmb$. This parametrization is identical to the one used in \cite{More_11}.
	
	\subsubsection{Satellites}
	
	Each halo can have an arbitrary number of satellite galaxies. We assume that the occupation is governed by a Poisson distribution with expectation value
	\begin{equation}\label{Nsat}
		N_\rms (M) = \int\limits_{L_{\rm th}}^\infty \Phi_\rms (L | M) \rmd L,
	\end{equation}
	where $L_{\rm th}$ is the luminosity threshold. We model the satellite CLF as a modified Schechter function:
	\begin{equation}\label{satCLF}
		\Phi_\rms (L | M) = \frac{\phi_\rms^* (M)}{L_\rms^*(M)} \left( \frac{L}{L_\rms^*(M)} \right)^{\alpha_\rms} \exp \left[ - \left( \frac{L}{L_\rms^* (M)} \right)^2 \right].
	\end{equation}
	In essence, the luminosities of satellites are drawn from a power-law with slope $\alpha_\rms$ with an exponential cut-off above a critical luminosity, $L_\rms^*(M)$. The luminosity cut-off, $L_\rms^*(M)$, is related to the characteristic luminosity of red, central galaxies in haloes of the same mass according to
	\begin{equation}
		L_\rms^*(M) = 0.562 \tilde L_{\rmc, \rmr} (M).
	\end{equation}
	This scaling is a good fit to the luminosity distribution of central and satellite galaxies as inferred from the SDSS galaxy group catalogue of \citet{Yang_09}. Finally, the normalization $\phi_\rms^* (M)$ is parametrized by
	\begin{equation}
		\log \left[ \phi_\rms^* (M) \right] = b_0 + b_1 \log M_{12} + b_2 (\log M_{12})^2.
	\end{equation}
	As described in detail in \S\ref{sec:mocks}, we assume that no satellite galaxy is brighter than its corresponding central. For the colour of satellite galaxies we choose the same parametrization as for the centrals. We find that with $f_{0, \rm sat} = 0.44$ and $\alpha_{f, \rm sat} = 0.14$ this is a very good approximation of the data reported in \cite{Yang_08} as a function of both halo mass and satellite luminosity. Also note that \cite{Yang_08} use a red and blue colour cut that is slightly different from the one used in \cite{More_11}. Since we are mainly interested in central galaxy colour, and ignore the colours of satellites candidates, this negligible difference only matters for the small amount of impurity in our central sample, as discussed in \S\ref{subsec:cylinders}. We make no attempt to fit $f_{0, \rm sat}$ and $\alpha_{f, \rm sat}$ with our data and therefore keep their values fixed. This leaves us with $4$ free parameters to describe the satellite CLF: $\alpha_s$, $b_0$, $b_1$, and $b_2$. This parametrization is identical to the one used by \cite{vdBosch_13} and \cite{Cacciato_13}.
	
	\subsubsection{Summary}
	
	\begin{table*}
		\begin{tabular}{l | l | l | l}
			Name & Description & Prior & Default \\
			\hline\hline
			$\log L_{0, \rmr/\rmb}$ & normalization of mass--luminosity relation for red/blue centrals & $[9.0, 10.5]$ & $9.99/9.55$\\
			$\log M_{1, \rmr/\rmb}$ & characteristic mass of mass--luminosity relation for red/blue centrals & $[10.0, 13.0]$ & $11.50/10.55$\\
			$\gamma_{1, \rmr/\rmb}$ & low-mass slope of mass--luminosity relation for red/blue centrals & $[0.0, 5.0]$ & $4.88/2.13$\\
			$\gamma_{2, \rmr/\rmb}$ & high-mass slope of mass--luminosity relation for red/blue centrals & $[0.0, 2.0]$ & $0.31/0.34$\\
			$\sigma_{\rmr/\rmb}$ & scatter in luminosity at fixed halo mass for red/blue centrals & $[0.1, 0.3]$ & $0.20/0.24$\\
			$f_0$ & red fraction of centrals at halo mass $M = 10^{12} \Msunh$ & $[0.0, 1.0]$ & $0.70$\\
			$\alpha_f$ & linear dependence of the red fraction of centrals on $\log \Mvir$ & $[-0.5, 1.0]$ & $0.15$\\
			$\alpha_s$ & low-luminosity slope of the satellite CLF & $[-1.5, -0.9]$ & $-1.18$\\
			$b_0$ & normalization of the satellite CLF & $[-1.5, 0.5]$ & $-0.766$\\
			$b_1$ & normalization of the satellite CLF & $[0.0, 2.0]$ & $1.008$\\
			$b_2$ & normalization of the satellite CLF & $[-0.5, 0.5]$ & $-0.094$\\
			\hline
		\end{tabular}
		\caption{Variables governing the occupation of dark matter haloes with galaxies. Note that the first $5$ parameters exist independently for red and blue centrals. Column (3) indicates the prior range used in our analysis, while column (4) lists the default values used for the construction of our mock data. See the text for details.}
		\label{tab:parameters}
	\end{table*}
	
	Table \ref{tab:parameters} lists all parameters that describe the galaxy occupation. It also lists their prior ranges used when fitting data and their default values used to create and analyse mock catalogues in \S\ref{sec:application_mocks}. These default values are taken from \cite{More_11} and \cite{Cacciato_09, Cacciato_13} for centrals and satellites, respectively. Note that we use a slightly different halo mass definition than \cite{More_11} or \cite{Cacciato_13}. Additonally, \cite{More_11} infer a luminosity scatter of $\sigma_{\rmr/\rmb} \sim 0.20 \ \mathrm{dex}$, whereas other studies \citep{Yang_08, Cacciato_13} suggest $\sigma \sim 0.15 \ \mathrm{dex}$. Since the main goal of this paper is to investigate our ability to recover input parameters from complex mock catalogues, these slight inconsistencies are not relevant.
	
	\subsection{Phase-space distributions}
	\label{subsec:phase-space_distributions}
	
	The CLF described above characterizes the abundance of central and satellite galaxies as a function of halo mass. We now describe our model for the phase-space coordinates of these galaxies with respect to their host halo.
	
	\subsubsection{Centrals}
	\label{sec:phase_cen}
	
	Throughout we assume that central galaxies reside at rest at the centre of their host haloes, which are assumed to be spherically symmetric. While it is known that centrals are not perfectly at rest relative to the centre-of-mass of their host halo, \citep[][]{vdBosch_05a, Behroozi_13, Guo_15a, Guo_15b, Guo_16, Ye_17}, this motion does not have a significant impact. The typical speed of a central is $\sim 20\%$ of the root-mean-square speed of its satellites \citep{Ye_17}. Thus, neglecting the motion of centrals will underestimate the central-satellite velocity dispersion by $\approx \sqrt{1 + 0.2^2} - 1 = 2\%$. This is smaller than our measurement uncertainties and other systematic contributions to our error budget.
	
	\subsubsection{Satellites}
	\label{sec:phase_sat}
	
	Satellite galaxies are assumed to follow a radial profile, $n_{\rm sat}(r)$, that is also spherically symmetric. Since we stack a large number of dark matter haloes with random orientations, this assumption of spherical symmetry is fair, despite the fact that individual haloes, and their satellite populations, are known to be aspherical \citep[e.g.,][]{Wang_08}. We follow \cite{vdBosch_04} and assume that the radial profile of satellite galaxies, as a function of the radial distance $r$ from the halo centre, is given by a generalized Navarro--Frenk--White (NFW) profile \citep{Navarro_96},
	\begin{equation}\label{nsatprof}
		n_{\rm sat} (r) \propto \left( \frac{r}{\mathcal{R} r_\rms} \right)^{-\gamma} \left( 1 + \frac{r}{\mathcal{R} r_\rms} \right)^{\gamma - 3}\,.
	\end{equation}
	Here $\mathcal{R}$ and $\gamma$ are free parameters and $r_\rms$ is the scale radius of the dark matter halo, which is related to the halo virial radius via the concentration parameter $c_{\rm vir} = \rvir / r_\rms$. By definition, no satellites are located outside the virial radius. Although we acknowledge that some galaxies physically associated with the halo can have orbits that take them outside the virial radius, for the purpose of this study we define them as interlopers. Throughout this work we consider three different radial profiles for satellites. The first has $\gamma = \mathcal{R} = 1$, which implies that satellites follow an NFW profile with the same concentration parameter as the dark matter; in other words, satellite galaxies are unbiased tracers of the dark matter within their host haloes. We also consider a profile with $\gamma = 1$ and $\mathcal{R} = 2.0$, for which the satellites still follow an NFW profile, but with a concentration parameter that is half that of the dark matter. The final profile that we consider has $\gamma = 0$ and $\mathcal{R} = 2.5$, which implies that the distribution of satellites has a core of constant number density, and a concentration parameter that is 2.5 times smaller than that of its dark matter. As we show in \S\ref{subsec:mocks_complex}, this latter profile is a good description of the distribution of subhaloes in dark matter simulations. Altogether, these profiles roughly bracket the range of observational constraints on the radial distribution of satellite galaxies in groups and clusters \citep[e.g.,][]{Carlberg_97, vdMarel_00, Lin_04, Yang_05, Chen_08, More_09b, Guo_12a, Cacciato_13}.
	
	In our analytical model, we assume that the velocities of satellites are isotropic, and obey the spherically symmetric Jeans equation \citep[][]{Binney_87}. For satellite galaxies with a radial profile given by Eq.~(\ref{nsatprof}), located in an NFW halo with virial velocity $\Vvir$ and concentration parameter $c_{\rm vir}$, this implies a velocity dispersion given by
	\begin{align}
		\sigma^2 (r | \Vvir, c_{\rm vir}) =& \frac{c_{\rm vir} \Vvir^2}{\mathcal{R}^2 g(c_{\rm vir})} \left( \frac{r}{\mathcal{R} r_\rms} \right)^\gamma\left( 1 + \frac{r}{\mathcal{R} r_\rms} \right)^{3 - \gamma} \nonumber\\
		& \int\limits_{r / r_\rms}^{\infty} \frac{g(y)\mathrm{d}y}{(y / \mathcal{R})^{\gamma + 2} (1 + y / \mathcal{R})^{3 - \gamma}}\,,
		\label{eq:Jeans}
	\end{align}
	where
	\begin{equation}
		g(y) = \ln (1 + y) - \frac{y}{1 + y}.
	\end{equation}

\begin{figure}
		\centering
		\includegraphics[width=\columnwidth]{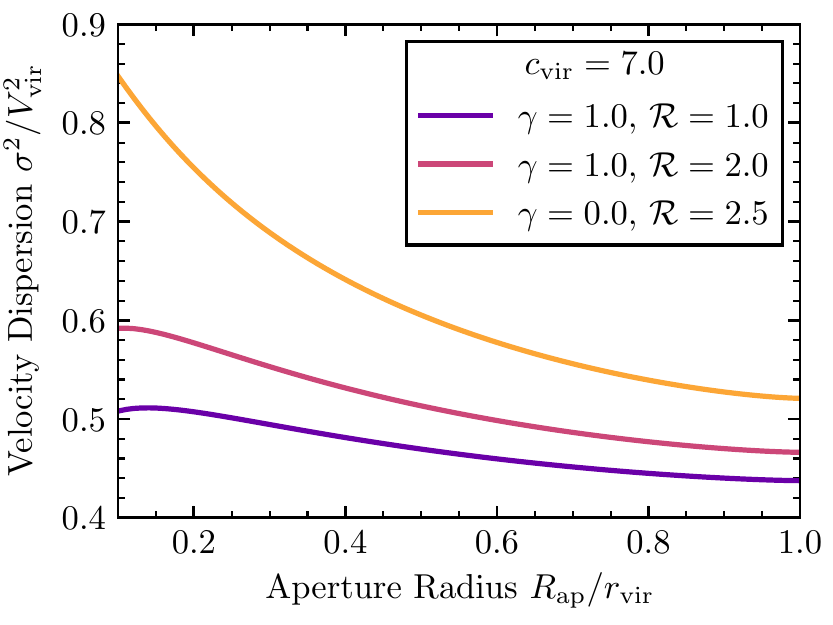}
		\caption{The average line-of-sight velocity dispersion of satellites within a projected aperture of radius $R_{\rm ap}$ implied by the spherically symmetric Jeans equation. Generally, the velocity dispersion decreases for larger aperture radii, as does the difference between the radial profiles. More extended radial profiles lead to a higher overall velocity dispersion.}
		\label{fig:profile_vs_dispersion}
	\end{figure}

	Figure \ref{fig:profile_vs_dispersion} shows how our choice of the radial profile affects the velocity dispersion. For each of our three radial profiles we plot the average velocity dispersion inside a fixed aperture centred on the halo. The concentration parameter is set to $c_{\rm vir} = 7$ for this illustration. By scaling the aperture radius by $\rvir$ and the dispersion by $V_{\rm vir}^2$ the relation becomes independent of halo mass. Note that less centrally concentrated satellite profiles imply a larger velocity dispersion. However, the differences decrease as the aperture encloses larger parts of the satellite population, declining from $\sim 50\%$ to $\sim 20\%$ when $R_{\rm ap} / r_{\rm vir}$ increases from $0.1$ to $1.0$.
	
	Our model for the velocities of satellite galaxies makes two crucial assumptions:
	isotropy and virial equilibrium. It is well known that dark matter subhaloes and satellite galaxies have kinematics that are slightly radially anisotropic \citep[e.g.,][]{Diemand_04, Wojtak_13}. However, velocity anisotropy mainly affects the {\it radial profile} of $\sigma^2(r)$; not the {\it total} velocity dispersion averaged over the entire satellite population (the latter appears in the virial theorem, which is independent of velocity anisotropy). Since we are measuring, and modelling, the velocity dispersion of satellite galaxies inside a relatively large aperture centred on the central galaxy, neglecting anisotropy only affects the resulting velocity dispersions by a few percent \citep[see][for details]{vdBosch_04, Mamon_05, Conroy_07}. The assumption of virial equilibrium is also subject to concern. After all,  dark matter haloes are constantly accreting matter, and new satellite galaxies, from their surroundings. Hence, some satellite galaxies will be on first infall, and their kinematics will not be in perfect virial equilibrium. As shown by \citet{Ye_17} ignoring the resulting time-dependence in the Jeans equations (which are based on the assumption of a steady-state), may result in small biases in the  inferred velocity dispersion. We address this issue in \S\ref{subsec:mocks_complex} by analysing mock catalogues where satellite galaxies are placed on subhaloes. Those tests suggest that deviations from the steady-state assumption do not cause a strong bias in our inferred model parameters.
	
	\section{Mock SDSS Catalogues}
	\label{sec:mocks}
	
	In order to test and validate our analytical model, we construct mock SDSS-like galaxy redshift surveys. By treating these mocks exactly as we treat the real data, we can test the accuracy to which we can recover the galaxy-dark matter connection, $P(\calH | \calG_1, \calG_2,...,\calG_n)$. 
	
	Our mocks are constructed using the SMDPL simulations \citep{Klypin_16}, which traces the dark matter distribution in a cosmology that is based on the \cite{Planck_14} cosmological parameters. It traces the dark matter distribution inside a cubic volume of $(400 \Mpch)^3$ using $3840^3$ particles, resulting in a particle mass of $1.0 \times 10^8 \Msunh$. We construct mock redshift surveys using the publicly available\footnote{http://yun.ucsc.edu/sims/SMDPL/trees/index.html} $z = 0.05$ \texttt{ROCKSTAR} halo catalogues \citep{Behroozi_13}, and use \texttt{halotools} \citep{Hearin_17a} to populate all dark matter haloes with $M > 3 \times 10^{10} \Msunh$ with mock galaxies according to our fiducial CLF model (see \S\ref{sec:CLF} and Table~\ref{tab:parameters}).
	
	For each host halo above this mass limit, we first decide on the colour of the central by drawing a random number between $0$ and $1$.  The central is assumed to be red (blue) if this random number is smaller (larger) than $f_\rmr(M)$. Next we draw the corresponding luminosity from $\Phi_{\rm c,r}(L|M)$, if red, or $\Phi_{\rm c,b}(L|M)$, if blue. Any central galaxy with a luminosity below a threshold of $10^9 \Lsunh$ is removed from the catalogue. The remaining central galaxies are given the position and velocity of their halo cores, which is defined as the region that encloses the innermost $10$ percent of the halo virial mass. These positions and velocities are calculated by \texttt{ROCKSTAR} as detailed in \citet{Behroozi_13}.
	
	Next we draw the number of satellite galaxies, under the assumption that $P(N_\rms|M)$ follows a Poisson distribution, whose mean is given by Eq.~(\ref{Nsat}). Throughout, we adopt a luminosity threshold of $10^9 \Lsunh$ for our mock galaxies. For each of the $N_\rms$ satellites in the halo in question we then draw a luminosity from the satellite CLF, $\Phi_\rms(L | M)$, given by equation (\ref{satCLF}) and a colour based on $f_{\rm r, sat}(M)$. After drawing satellites from the above CLF we remove all satellites that are brighter than their respective centrals\footnote{By removing bright satellites the satellite occupation is slightly lowered and has a small dependence on the actual central luminosity. However, we find this decrease and its impact on satellite kinematics to be negligible. This is because it is generally unlikely that any given satellite is brighter than the central in CLF models \citep{Skibba_11, Lange_18a}. In our fiducial mock, this only removes $1\%$ of all satellites.}. For the positions and velocities we assume the analytical model for the phase-space coordinates of satellite galaxies described in \S\ref{sec:phase_sat} above; we draw the positions from a spherical distribution with radial profile $n_\rms(r)$, given by Eq.~(\ref{nsatprof}), and one-dimensional velocities from a Gaussian distribution with dispersion $\sigma(r)$, given by Eq.~(\ref{eq:Jeans}). Both the positions and the velocities are with respect to the core of the host halo.\footnote{A more realistic approach would be to express all phase-space positions of satellites with respect to the bulk of the halo. However, this information is not available in the public halo catalogues. As discussed previously, the impact of this simplification on the velocity dispersion should be negligible.}
	
	After populating the dark matter haloes with mock galaxies, we place a virtual observer inside the simulation box. Both the position and the orientation are chosen randomly. We then proceed to calculate angular coordinates, redshifts and apparent magnitudes for all galaxies within $z < 0.15$. If necessary, the simulation box is periodically repeated out to that redshift. The apparent magnitudes include $K$ and evolution correction effects. Next, an apparent magnitude cut of $m_r \leq 17.6$ is applied, after which we add redshift space distortions to the cosmological redshifts. In addition, we also add a random scatter in redshift of the order of $15\kms / c$ to simulate redshift uncertainties in SDSS \citep{Guo_15b}. Subsequently, we apply the SDSS DR7 survey mask\footnote{\url{http://sdss.physics.nyu.edu/lss/dr72}}, rejecting galaxies outside of the survey window or inside of masked regions.
	
	Finally, we simulate the effect of spectroscopic incompleteness, particularly fibre collisions. In the SDSS, on a single plate spectroscopic fibres cannot be placed simultaneously for objects separated by less than $55''$ \citep{Blanton_03a}. However, some galaxies are observed with multiple plates, allowing spectroscopic redshifts even for close pairs. This fibre collision effect leads to some objects missing spectroscopic redshifts. We follow the common practice of assigning such `fibre-collided' galaxies the redshift of its nearest neighbour \citep[][]{Zehavi_05}. To simulate this type of incompleteness, we first construct a maximal `decollided' set of galaxy targets \citep{Blanton_03a}. Within this set of decollided galaxies, no two targets are within $55''$ of each other. The remaining galaxies are potentially collided targets that would not be assigned a redshift in SDSS if they would not lie in regions of overlapping spectroscopic plates. We then randomly choose $65\%$ of those galaxies to be missing a spectroscopic redshift. This value is very close to what we observe in the New York University Value-Added Galaxy Catalog \citep[VAGC;][]{Blanton_05}. See appendix \ref{sec:tiling} for more details. Additionally, we randomly remove an additional $1\%$ of all spectroscopic redshifts to simulate other failure modes \citep{Blanton_03a}. While this algorithm reproduces the most salient features of fibre collisions in SDSS, it misses less important details like the tiling algorithm and the according correlation of fibre collision probability with large-scale structure \citep{Blanton_03a}. We test the impact of this correlation on our observables in appendix \ref{sec:tiling} but find it negligible. Thus, the fibre collision algorithm simulated here suffices for our mock simulations.
	
	\section{Observations}
	\label{sec:observations}
	
	In this section we first describe the detailed criteria used to select primary (assumed to be central) and secondary (assumed to be satellite) galaxies from a galaxy redshift survey. We then describe how we use these galaxies to measure the host- and satellite-weighted velocity dispersions and other observables that we use to model the galaxy-dark matter connection. 
	
	Throughout this study, we only use galaxies in a volume-limited sample specified by $0.02 \leq z \leq 0.067$, which, for our adopted apparent magnitude limit of $m_r = 17.6$, is complete down to a $r$-band luminosity of $10^{9.5} h^{-2}\Lsun$. The main reason for using a volume-limited sample is to facilitate constraining the scatter in the galaxy-halo connection via a comparison of host- and satellite-weighted velocity dispersions \citep[see][for details]{More_09a}. In addition, if one were to use a flux-limited sample, one also needs to model potential luminosity segregation, since haloes at a lower redshift, are now sampled down to lower satellite luminosity. Mainly for these two reasons, we accept the price of a reduced dynamic range and a reduced signal-to-noise that comes with using only a volume-limited subsample of the full SDSS data. We hope to extent our methodology to flux-limited samples in future work.
	
	\subsection{Sample selection criteria}
	\label{subsec:cylinders}
	
	We use cylindrical isolation criteria to select samples of centrals and satellites. Due to interlopers and other impurities, not every central (satellite) thus selected is indeed a central (satellite). We therefore refer to galaxies that are selected as centrals and satellites as primaries and secondaries, respectively.
	
	A galaxy is considered a primary if it is at least $f_\rmh$ times brighter than any other galaxy within a cylindrical volume specified by $R_\rmp < R_\rmh$ and $|\dv | < (\Delta v)_\rmh$. Here $R_\rmp$ is the separation projected on the sky at the distance of the primary candidate and $\dv$ is the line-of-sight (los) velocity difference (Eq.~[\ref{eq:dv}]). We apply this isolation criterion in a rank-ordered fashion, starting with the brightest galaxy. If a galaxy is selected as a primary, we remove all other galaxies within its $[R_\rmh,(\Delta v)_\rmh]$ cylinder from the list of potential primary candidates. Around each primary galaxy thus selected, secondaries are defined as those galaxies that are at least $f_\rms$ times fainter than their primary and located within a cylindrical volume with $R_\rmp < R_\rms$ and $|\dv| < (\Delta v)_\rms$. After having identified primaries, we remove those without a spectroscopic redshift. Furthermore, to avoid complications due to survey edge effects, we remove all primaries for which more than $20\%$ of a ring with radius $R_\rmh$ centred on the primary falls outside of the survey window or inside of the mask.
	
	In total, the selection of primaries and secondaries thus depends on six free parameters: $R_\rmh$, $(\Delta v)_\rmh$ and $f_\rmh$ to specify the population of primaries, and $R_\rms$, $(\Delta v)_\rms$ and $f_\rms$ to specify the secondaries.  These parameters determine the completeness and purity of the sample, both in terms of primaries and secondaries. They also determine the interloper fraction, which is defined as the fraction of secondaries that are not satellite galaxies within the same halo as the corresponding primary. Minimizing the number of interlopers requires sufficiently small $R_\rms$ and $(\Delta v)_\rms$, while maximizing the purity of the primaries (i.e., minimizing the number of satellites that are erroneously identified as centrals) requires large $R_\rmh$, $(\Delta v)_\rmh$ and $f_\rmh$. Of course, each of these restrictions dramatically reduces the completeness of the samples, thereby making the measurements more noisy.
	
	Most previous studies of satellite kinematics have adopted extremely conservative selection criteria, with both $f_\rmh$ and $f_\rms$ significantly larger than unity \citep[e.g.,][]{Zaritsky_93, Zaritsky_97, McKay_02, Brainerd_03, Prada_03, Conroy_07, Norberg_08, Wojtak_13}. In addition, all these studies adopted fixed values for the selection criterion parameters, independent of the luminosity of the galaxy under consideration. Since brighter centrals on average reside in more extended haloes,  \citet{vdBosch_04} advocated an aperture which scales with the virial radius of the primary's halo. They used an iterative criterion whereby the cylindrical aperture scales with the velocity dispersion around the primary obtained in the previous iteration. By using such an adaptive, iterative selection criterion, they were able to increase the number of secondaries by almost an order of magnitude (using $f_\rmh = f_\rms = 1$), while drastically reducing the actual interloper fraction.
	
	The same iterative method was later used in \cite{More_09b} and \cite{More_11}, who found that the satellite velocity dispersion around red primaries scales with luminosity $L$ as
	\begin{equation}
		\log \sigma_{200} \equiv \log \left( \frac{\sigma}{200 \kms} \right) = a + b \log L_{10} + c \log^2 L_{10},
		\label{eq:sigma_cylinder}
	\end{equation}
	where $L_{10} = L / 10^{10} \Lsunh$ and $a = -0.07$, $b = 0.38$ and $c = 0.29$\footnote{Note that Eqs. (11) in \cite{More_09b} and (3) in \cite{More_11} are incorrect, and inconsistent with the values of $a$, $b$ and $c$ quoted in those papers. The equation presented here is the one actually used by \cite{More_11} for their LR sample. (Surhud More, {\it priv. comm.})}. We base our selection of primaries on this scaling relation. We follow \citet{vdBosch_04} and \cite{More_11} by setting $f_\rmh = f_\rms = 1$, and scale the dimensions of the cylinder used to select primaries as $R_\rmh = 0.5 \sigma_{200} \Mpch$ and $(\Delta v)_\rmh = 1000 \sigma_{200} \kms$. This value for $R_\rmh$ roughly corresponds to $1.25$ times the virial radius. We have experimented with other cylinder sizes for the isolation criterion, but typically find that increasing completeness results in a larger contamination (i.e., reduced purity). In the end, we decided on values that result in at least $\sim 99\%$ purity in centrals for the primary sample.
	
	Note that, contrary to the approach adopted here, \cite{More_09b, More_11} did not remove galaxies located inside the larger cylinder of a brighter galaxy from the list of potential primary candidates. On the other hand, they used $R_\rmh = 0.8 \sigma_{200} \Mpch$ instead of $0.5 \sigma_{200} \Mpch$. Coincidentally, we find that both strategies result in very similar completeness and purity. Contrary to \cite{More_11}, we choose the same cylinder sizes for red and blue primaries. This is done primarily to not bias our selection towards either blue or red primaries. In order to avoid excessive values for $\sigma_{200}$, which can occur for fibre-collided galaxies that are accidentally assigned a too high redshift, we limit $\sigma_{200}$ to a maximum value of $5$, i.e. $\sigma \leq 1000 \kms$.
	
	The criterion for secondaries is that they must have a spectroscopic redshift (i.e., they can not be fiber-collided galaxies) and that they must lie in a cylinder centred on a primary with $R_\rms = 0.15 \sigma_{200} \Mpch$ and $(\Delta v)_\rms = 4000 \kms$. This value for $R_\rms$ roughly corresponds to $0.375$ times the virial radius, while the value for $(\Delta v)_\rms$ is large enough to include the vast majority of all satellites, even in massive clusters. Note that we do not scale this parameter with $\sigma_{200}$; although this implies an increasing fraction of interlopers with decreasing central luminosity, these interlopers are easily identified as such. Furthermore, having a pure sample of interlopers at large velocity offsets allows us to better estimate interloper contamination at lower velocities (see \S~\ref{subsec:interloper_rejection}).
	
	Contrary to the selection of primaries, for the selection of secondaries, we permit the cylinder sizes to differ as a function of galaxy colour: for red primaries we use equation~(\ref{eq:sigma_cylinder}) with the parameters listed above, while for blue primaries we use $a = -0.19$, $b = 0.46$ and $c = -0.16$. The different cylinder dimensions for blue centrals account for their different average dark matter halo masses \citep[see][]{More_11}. Finally, in order to alleviate the impact of  imperfect correction for fibre-collisions (see \S\ref{subsec:fibre_collisions}), we exclude secondaries that have a projected distance to their primaries of $R_\rmp \leq R_\rmc \equiv 60 \kpch$. 
	\begin{figure*}
		\centering
		\includegraphics[width=\textwidth]{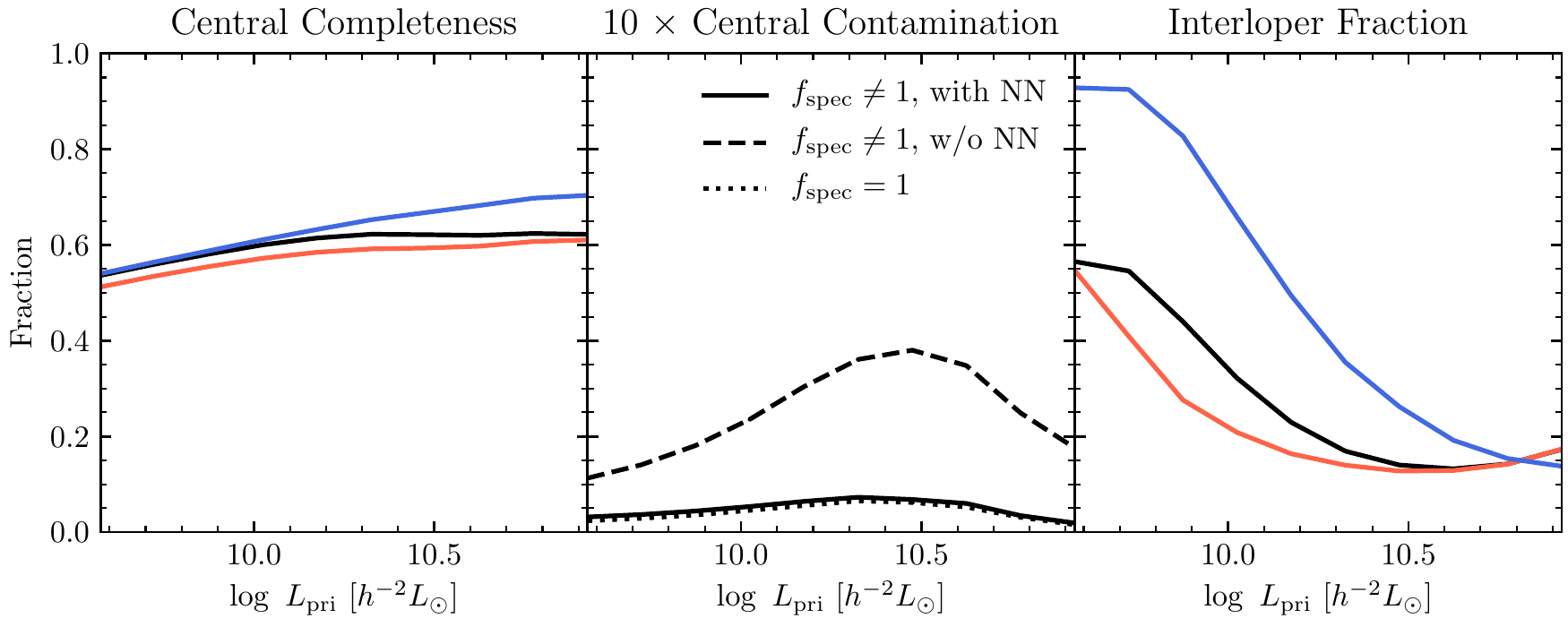}
		\caption{This plot shows the efficiency of our selection criterion as inferred from mock catalogues built with the default model. We display the central completeness, the fraction of centrals in the survey volume selected as primaries (left panel), the central contamination, the probability of a primary to not be a central (middle panel), and the interloper fraction, which is defined as the fraction of secondaries not residing in the same halo as the primary (right panel). Note that we also include primaries without any secondaries when calculating central completeness and purity. All probability are displayed as a function of primary luminosity. Additionally, we show them for red primaries (red), blue primaries (blue) and all (black). The dotted line in the middle panel shows the contamination if there was $100\%$ spectroscopic completeness, e.g. no fibre collisions. On the other hand, the dashed line is the same fraction in a spectroscopic survey with incompleteness where fibre collided galaxies are not assigned the redshift of the nearest neigbour (NN) and instead removed before applying the selection criterion. Note that the central contamination is multiplied by $10$ to fit into the same range as the other ratios.}
		\label{fig:efficiency}
	\end{figure*}
	
	We explore the efficiency of our selection criterion in Figure \ref{fig:efficiency}. All quantities are inferred from SDSS-like mock catalogues, constructed using the method described in \S\ref{sec:mocks} in which satellites follow an NFW profile. The left-hand panel displays the central completeness, defined as the fraction of centrals that are ultimately selected as primaries. Ideally, we want this to be close to unity. Our selection criteria achieve a central completeness of around $\sim 60\%$ for all luminosities. Note that a significant fraction of incompleteness at the high luminosity end, especially for red centrals, is caused by spectroscopic incompleteness due to fibre collisions. The increased occurrence of fibre collisions is expected due to the stronger clustering of red or more bright galaxies. The solid curve in the middle panel shows that the central contamination, defined as the fraction of primaries that are not centrals, is extremely low, less than $1\%$ for all primary luminosities. The dotted line shows the same quantity in a mock with $100\%$ spectroscopic completeness (i.e., no fibre collisions), which results in a very similar contamination fraction. Hence, central contamination is not dominated by  catastrophic failures in the nearest neighbour redshift assignment of fibre-collided galaxies, which can happen if the fibre-collided central of a halo is assigned the redshift of a fore- or background galaxy. Finally, the dashed line shows the central contamination if one removes fibre collided galaxies {\it before} identifying primaries. This results in a contamination fraction that is roughly 5 times larger. This demonstrates that assigning fibre-collided galaxies the redshift of its nearest neighbour is reasonably accurate \citep[see also][]{Zehavi_11}, and that one should only remove fibre collided galaxies {\it after} the primaries have been identified.
	
	To close this subsection, we note that in our mocks centrals are always the brightest galaxies in their haloes. This is contrary to various studies showing that in a small fraction of haloes a satellite will be the brightest galaxy \citep{Skibba_11, Hikage_13, Wang_14, Hoshino_15, Lange_18a}. We have tested that, if that is the case, our sample of primaries should be very pure in brightest halo galaxies (hereafter BHGs), rather than central galaxies. One might worry that some of our supposed centrals being satellites would systematically bias our inferences. For example, one might naively expect that the measured velocity dispersion is $\sqrt{2}$ times higher when measured around satellites. However, in \cite{Lange_18a}, we have shown that ignoring this complication does not have a significant impact on our inferences \citep[see also][]{Skibba_11}.
	
	\subsection{Interloper rejection}
	\label{subsec:interloper_rejection}
	
	One of the main challenges in measuring satellite kinematics is that not all secondaries reside in the same halo as their primary. Such interlopers have a very different velocity distribution than true satellites, making it crucial that one corrects for this interloper contamination. The right-hand panel of Figure \ref{fig:efficiency} shows the fraction of interlopers among secondaries. For this figure we only counted secondaries with $|\dv| < 600 \sigma_{200} \kms$. Secondaries with larger velocity differences are trivially identified as interlopers. We first note the extremely high interloper fraction for low-luminosity blue primaries. That is mainly due to the fact that those galaxies have very few same-halo satellites in our model. However, for the rest of the primaries, the selection of secondaries results in interloper fractions as low as $\sim 15\%$.
	
	The $\dv$ distribution of secondaries consists of a satellite component centred on $\dv = 0$ and an interloper component that is close to flat. Most previous studies modelled the $\dv$ distribution of satellites as a single Gaussian and the interloper distribution as flat \citep[][]{McKay_02, Prada_03, Brainerd_03, vdBosch_04, Conroy_07}. In this case, the velocity dispersion of the best-fit Gaussian is the estimate for the dispersion of same-halo satellites. However, \cite{Becker_07} showed that a single Gaussian fit significantly underestimates the dispersion of same-halo satellites. \cite{More_09b} advocated using a fit to a double-Gaussian plus a flat background model to estimate the velocity dispersion of satellites. Still, contrary to the claims by \cite{More_09b}, we also find this dispersion estimate to be biased. Similar to what was estimated by \cite{Becker_07}, we find that it under-predicts the dispersion by roughly $3\%$. We have also tested the bi-weight estimator \citep{Beers_90}, but encountered similar biases. Here, we use a different method to reject interlopers. While this technique does not significantly improve the accuracy in the velocity dispersion estimate (i.e., it does not significantly reduce the bias), it does improve the precision of the estimator slightly. We later correct for the bias numerically using a large number of mock observations for which the true velocity dispersions are known.
	
	The main idea of our interloper rejection is to use the distribution of galaxies in the $\dv$--$R_\rmp$ plane instead of just using velocity information \citep[also see][]{Klypin_09}. Thus, we model the total phase-space distribution of secondaries around primaries as the sum of an interloper component $P_{\rm int} (\dv, R_\rmp)$ and a model for the phase-space distribution of satellites around centrals $P_{\rm sat}(\dv, R_\rmp)$. Thus,
	\begin{equation}
		P_{\rm tot} (\dv, R_\rmp) = f_{\rm int} P_{\rm int} (\dv, R_\rmp) + (1 - f_{\rm int}) P_{\rm sat} (\dv, R_\rmp).
	\end{equation}
	For the interloper model we assume a flat distribution in $\dv$ and a constant projected number density, i.e., 
	\begin{equation}
		P_{\rm int} (\dv, R_\rmp) \propto R_\rmp.
	\end{equation}
	Based on tests with our mock catalogues, we find the assumption regarding the $R_\rmp$  distribution to be fairly accurate \citep[see also][]{Wojtak_13, Zheng_16}. On the other hand, similar to \cite{vdBosch_04} and \cite{More_09b}, we find an increased number of interlopers at $\dv \sim 0$. However, this only has a minor impact on the estimate of the satellite velocity dispersion, and this offset is accounted for in our bias treatment described in \S\ref{sec:analysis}.
	
	Regarding the model for the satellites, we assume that they have a spherically symmetric density profile and obey the Jeans equation. See \S\ref{subsec:phase-space_distributions} for more details. Furthermore, we assume that the host halo masses of satellites are drawn from a log-normal distribution and that the concentration parameter $c_{\rm vir}$ at each host halo mass is the median concentration for that mass extracted from SMDPL.
	
	In each of the $20$ bins of colour and luminosity of primaries, we fit the $\dv$--$R_\rmp$ distribution of secondaries with a $3$-parameter model: the interloper fraction $f_{\rm int}$, the median halo mass $\tilde{M}$ and the spread in halo masses $\sigma_{\log M}$. We determine the fit that maximizes the likelihood,
	\begin{equation}\label{likemodel}
		\mathcal{L} \propto \prod_{\rm i}^{n} P_{\rm tot}^{w_i} (\dvi, r_{\rmp, i} | f_{\rm int}, \tilde{M}, \sigma_M),
	\end{equation}
	where the product runs over all secondaries in the sample and $w$ denotes an optional weight assigned to each secondary. The choice of weights is discussed in the next two sections.  Afterwards, a membership probability $p_{\rm mem}$ is assigned to each secondary,
	\begin{equation}\label{pmem}
		p_{\rm mem} (\dv, R_\rmp) = \frac{(1 - f_{\rm int}) P_{\rm sat} (\dv, R_\rmp)}{P_{\rm tot} (\dv, R_\rmp)},
	\end{equation}
	based upon the best-fit model.
	\begin{figure}
		\centering
		\includegraphics[width=\columnwidth]{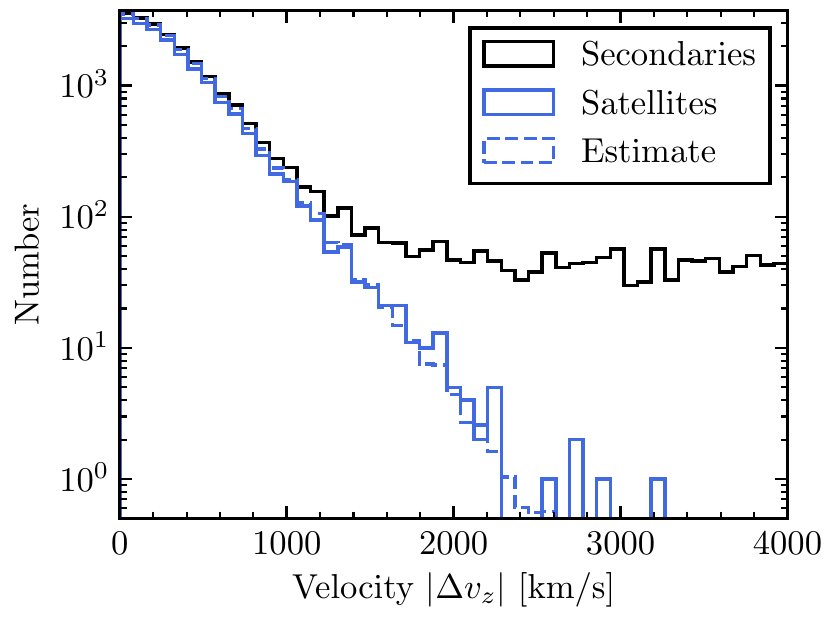}
		\caption{The distribution of line-of-sight velocities of secondaries around red primaries with $10.70 < \log L < 10.85$. We show all secondaries (black, solid), same-halo secondaries (blue, solid) and our estimate for the latter (blue, dashed) using the distribution of all secondaries. See the text for details.}
		\label{fig:interloper_fitting}
	\end{figure}
	
	Those membership probabilities can then be used in a straightforward manner to estimate the velocity dispersion of same-halo satellites, excluding interlopers. An example is shown in Figure \ref{fig:interloper_fitting}. For this plot we have populated SMDPL with galaxies and used the distant observer approximation. No fibre collision effects were included and the weights in Eq.~(\ref{likemodel}) were set to unity. We show the line-of-sight velocity difference of secondaries around red centrals with $10.70 < \log (L/h^{-2}\Lsun) < 10.85$. The black solid line is the distribution of all secondaries, whereas the blue solid line only shows same-halo satellites. The blue dashed line is our estimate for the latter by weighting each secondary by its membership probability.
	
	Finally, while the $3$ best-fit parameters of the fit contain information about average halo masses and their spread, the true distribution in halo masses is more complicated and will be discussed below. Thus, those best-fit parameters are disregarded and not used for further analysis. In other words, the model for $p_{\rm mem} (\dv, R_\rmp)$ discussed here is solely used to estimate membership probabilities.
	
	\subsection{Fibre collision correction}
	\label{subsec:fibre_collisions}
	
	When running our selection criterion to identify primaries and secondaries, we assign each galaxy missing a spectroscopic redshift that of its closest neighbour \citep[see][]{Blanton_05, Zehavi_05}. However, we stress that in our subsequent analysis only primary-secondary pairs with spectroscopic redshifts for both are used. Additionally, we assign each galaxy a spectroscopic weight $w_{\rm spec}$ to counteract the effect of fibre collisions. For each galaxy we count the number of galaxies inside a radius of $55''$ around it. We then determine for the entire survey what fraction $f_{\rm spec}$ of galaxies with exactly that number of neighbours within $55''$ have been assigned a spectroscopic redshift. The weight designated to the galaxy is then simply $w_\rms = 1/f_{\rm spec}$. As discussed below, this correction works very well but not perfectly below the fibre collision scale. Thus, in order to be conservative, we only consider secondaries separated from the primary by at least $R_\rmc = 60 \kpch$, roughly the fibre collision scale at the maximum redshift of our analysis. This removes less than $10\%$ of secondaries around primaries with $L_{\rm pri} > 10^{10} \Lsunh$.
	\begin{figure}
		\centering
		\includegraphics[width=\columnwidth]{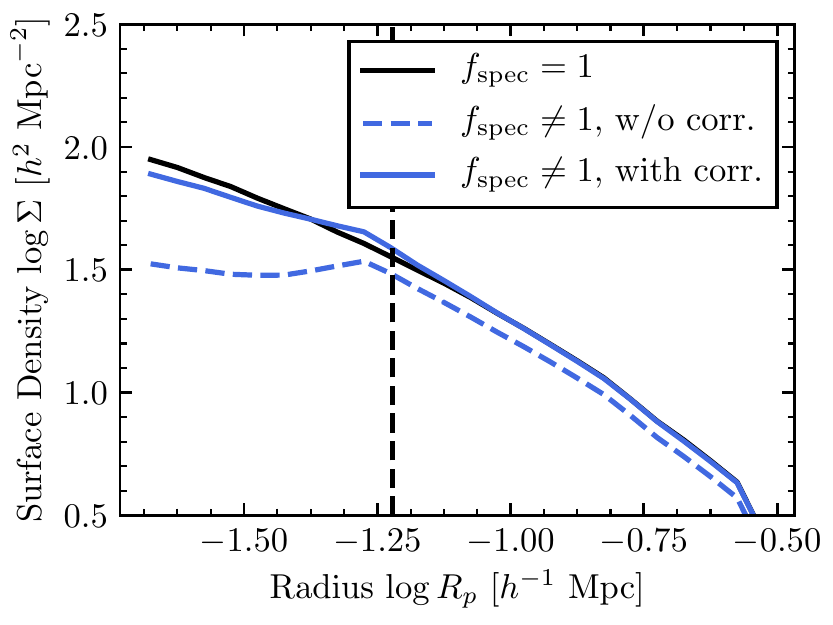}
		\caption{The surface density of same-halo secondaries around primaries with $10.6 < \log L / (\Lsunh) < 10.8$. The surface densities have been averaged over $100$ mock SDSS-like catalogues. We show the true surface density for surveys without fibre collisions (black), surveys having fibre collisions but no correction (blue, dashed) and surveys with fibre collisions and a correction applied to it (blue, solid). Secondaries at separations less than $60 h^{-1} \mathrm{kpc}$, as shown by the black dashed line, are excluded from our analysis. See the text for details.}
		\label{fig:fiber_collision_profile}
	\end{figure}
	
	We test the fidelity of our correction using mock catalogues. Particularly, for this test we assume that satellites follow the dark matter density distribution. In Figure \ref{fig:fiber_collision_profile} we show the surface density of secondaries around primaries with $10.6 < \log L / (\Lsunh) < 10.8$ in $100$ simulated SDSS-like surveys. Since we know true halo associations, we only show same-halo satellites. The black line is the idealized case of a spectroscopic survey without fibre collisions. The blue dashed line is for a survey with fibre collisions where we only consider galaxies with spectroscopic redshifts. We see an overall decrease in the surface density. Furthermore, there is a flattening of the profile around the fibre collision scale which is at $\sim 60\kpch$ at the maximum redshift of our survey and shown by the black dashed line. Finally, the blue solid line is the surface density estimate if the weight of each secondary is set to the product of $w_\rms$ of itself and its primary and the weight of each host is the primary weight. We see that this correction rectifies the overall decrease and also most of the effects below the fibre collision scale.
	\begin{figure}
		\centering
		\includegraphics[width=\columnwidth]{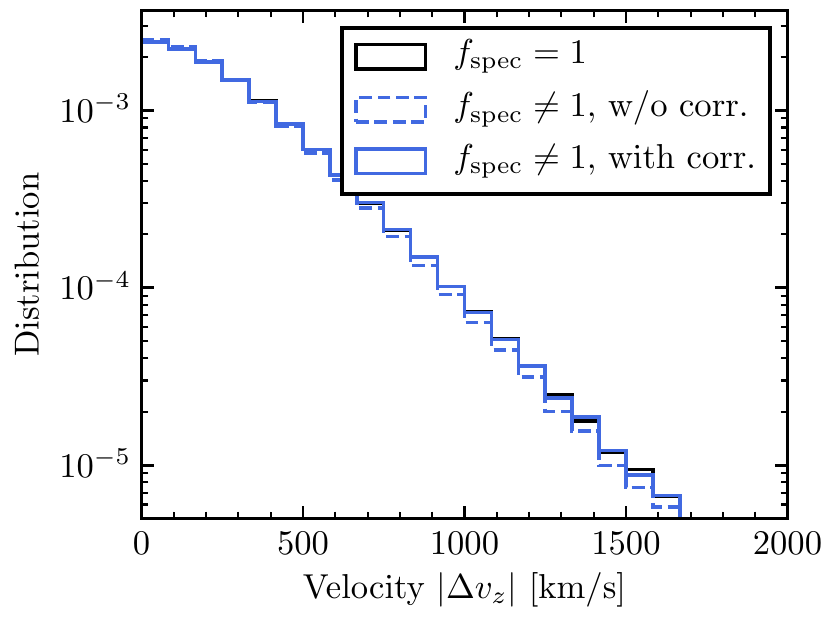}
		\caption{The line-of-sight velocity distribution of secondaries around primaries with $10.6 < \log L / (\Lsunh) < 10.8$. The primaries and secondaries are the same as in Figure \ref{fig:fiber_collision_profile}, except for excluding those secondaries with $R_\rmp < 60 h^{-1} \mathrm{kpc}$. As in Figure \ref{fig:fiber_collision_profile}, we show the distributions for surveys without fibre collisions (black), surveys having fibre collisions and no correction (blue, dashed) and surveys with collisions and a correction (blue, solid).}
		\label{fig:fiber_collision_vz}
	\end{figure}
	
	In addition to biasing the observed radial profile, fibre collisions also slightly alter the average host halo masses of observed primaries. The reason is that more massive haloes have more satellites and thus a higher probability to have a fibre collision between a primary and a secondary. Since we exclude primaries with fibre collisions, this biases our halo mass distribution at a fixed primary luminosity to lower masses. This effect is shown in Figure \ref{fig:fiber_collision_vz} where we show the line-of-sight velocities of the same secondaries as in Figure \ref{fig:fiber_collision_profile}, excluding those below $60 h^{-1} \mathrm{kpc}$. The measured velocity dispersion of secondaries if not correcting for fibre collisions is $\sim 375 \mathrm{km/s}$, whereas it is $\sim 388 \mathrm{km/s}$ without fibre collisions or with our correction applied. Thus not correcting for fibre collisions results in a velocity dispersion that is artificially lowered by around $5 \%$. For other luminosity bins, we find results that are qualitatively and quantitatively similar.
	
	Finally, we note that the corrections discussed above have not been applied in most other satellite kinematic studies \citep{McKay_02, Prada_03, Brainerd_03, Conroy_07, More_09b, More_11}. Based upon Figure \ref{fig:fiber_collision_profile}, it is apparent that neglecting these corrections can lead analysts to infer biased satellite density profiles and velocity dispersions. Particularly, this casts some doubt on the result of \cite{More_09b} that satellites are strongly anti-biased with respect to dark matter. This can be very significant in kinematical analyses because satellites have velocities that reflect their relative positions within the potential well of the group, and such biases in the spatial distributions of galaxies propagate into biases in the velocity distributions of galaxies (see the our previous discussion of the Jeans Equation as well as the discussion in \S\ref{sec:application_mocks} for more details on the effect of the spatial distribution of galaxies on observables). Specifically, the more extended radial profile assumed by \cite{More_09b, More_11} could lead to systematically underestimated halo masses. In addition to this radial profile bias, we have shown that there is a further bias in the velocity dispersion of a few percent if fibre collisions are not corrected. This can lead to an additional underestimation of average halo masses. Our analysis corrects for each of these biases.
	
	\subsection{Definition of Observables}
	\label{subsec:observables}
	
	We now describe the various observables that we use in our analysis of satellite kinematics. In addition to the host- and satellite-weighted velocity dispersions around red and blue primaries, this includes the abundances of secondaries around red and blue primaries, the fraction of primaries that are red, and the overall number density of galaxies. The latter is important to constrain the overall normalization of the CLF. For example, the best-fit model of \cite{More_11}, who did not include the overall abundance of galaxies as a constraint, implies\footnote{We have computed this using \texttt{Colossus} \citep{Diemer_15}, using the mass function of \cite{Tinker_08} and the cosmology adopted by \cite{More_11}.} a number density of galaxies brighter than  $10^{9.5} \Lsunh$ that is $\sim 2.9 \times 10^{-2} h^3 \mathrm{Mpc}^{-3}$. However, the observed number density is only $\sim 1.6 \times 10^{-2} h^3 \mathrm{Mpc}^{-3}$ \citep{Zehavi_11, Guo_15b}.
	
	Given our volume-limited sample, the number density in a luminosity bin defined by $L_1$ and $L_2$ is
	\begin{equation}
		n_{\rm gal} = \frac{\sum w_{\rms, i}}{\frac{\Omega_{\rm SDSS}}{3} 
			\left[ d_{\rm com}^3(z_{\rm max}) - d_{\rm com}^3(z_{\rm min})\right]}\,.
	\end{equation}
	Here $d_{\rm com}(z)$ is the comoving distance out to redshift $z$, the sum goes over all galaxies with $L_1 \leq L \leq L_2$, the angular sky coverage of our sample is $\Omega_{\rm SDSS} = 2.27$ steradians, and $w_{\mathrm{s}, i}$ is the weight of galaxy $i$ used to compensate for spectroscopic incompleteness, as discussed above. Throughout this paper we use 10 luminosity bins of 0.15 dex width covering the range $\log[L/(h^{-2}\Lsun)] = [9.5, 11.0]$. 
	
	The fraction of red primaries in each luminosity bin, $f_{\rm pri, r}$, is calculated according to
	\begin{equation}
		f_{\rm pri, r} = \frac{\sum\limits_{\rm red \ primaries}w_{\rms, i}}{\sum\limits_{\rm all \ primaries} w_{\rms, i}}\,.
	\end{equation}
	
	An important constraint on the CLF is the number of satellite galaxies per central. This is related to the average number of same-halo secondaries per primary, which we estimate as
	\begin{equation}
		\langle N_\rms \rangle = \frac{\sum\limits_{\rm secondaries} p_{\mathrm{mem}, i} w_{\mathrm{sw}, i}}{\sum\limits_{\rm primaries} w_{\rms, i}}\,.
	\end{equation}
	Here the sum in the numerator goes over all secondaries having primaries in that particular bin of luminosity and colour, while the sum in the denominator goes over all primaries of that bin. $p_{\mathrm{mem}, i}$ is the membership probability of secondary $i$, obtained from fitting the $\dv$--$R_\rmp$ distribution in each primary bin using the method described in \S\ref{subsec:interloper_rejection}, and $w_{\rm sw}$ is the weight assigned to each secondary, which is equal to the product of the spectroscopic incompleteness weights of the secondary and the corresponding primary, i.e., 
	\begin{equation}
		w_{\rm sw} = w_{\rms, \rmp} w_{\rms, \rms},
	\end{equation}

	Following \cite{More_09a}, we quantify the kinematics of satellite galaxies using both the satellite- and host-weighted velocity dispersions, $\sigma_{\rm sw}$ and $\sigma_{\rm hw}$, respectively. The former is computed by giving equal weight to each central-satellite pair, while the latter assigns equal weight to each central.  Since more massive haloes, on average, contain more satellites, the ratio $\sigma_{\rm sw}/\sigma_{\rm hw}$ contains information regarding the scatter in the galaxy-halo relation. In particular, $\sigma_{\rm sw}/\sigma_{\rm hw}$ increases with increasing scatter. We estimate the satellite-weighted velocity dispersion as
	\begin{equation}
		\sigma_{\rm sw}^2 = \frac{\sum\limits_{\rm secondaries} p_{\mathrm{mem}, i} w_{\mathrm{sw}, i} \dvi^2}{\sum\limits_{\rm secondaries} p_{\mathrm{mem}, i} w_{\mathrm{sw}, i}} - \sigma^2_{\rm err}\,.
		\label{eq:sigma_sw}
	\end{equation}
	Here $\sigma_{\rm err} = \sqrt{2} \times 15 \kms$ is the error on the velocity difference between primary and secondary arising from the redshift uncertainties in SDSS \citep[e.g.,][]{Guo_15b}.
	In the case of host-weighting, we use the same equation, but with $w_{\rm sw}$  replaced by
	a new weight, 
	\begin{equation}
		w_{\rm hw} = \frac{w_{\rms, \rmp} w_{\rm \rms, \rms}}{N_{\rm scd}}\, , 
	\end{equation}
	where $N_{\rm scd}$ is the number of secondaries hosted by each primary. Note that, when using these weights, we also use new membership probabilities, $p_{{\rm mem}, i}$, that are appropriate for calculating host-weighted quantities. These are computed using equation (\ref{pmem}), but with the weights in equation (\ref{likemodel}) replaced by $w_{\rm hw}$.
	
	Finally, we note that we only use the resulting $\langle N_\rms \rangle$, $\sigma_{\rm sw}^2$ and $\sigma_{\rm hw}^2$ for primary bins with at least an estimated $10$ satellites, i.e., with $\sum p_{\rm mem} > 10$.
	
	\section{Analytical Model for Observables}
	\label{sec:model}
	
	Ideally one would want to forward-model all relevant observational effects to compute the expectation values of the observables for a given set of model parameters. However, as discussed in \S\ref{sec:methodology}, this is computationally infeasible given the large number of likelihood evaluations needed to estimate parameter uncertainties. Instead, we estimate the expectation values for a given set of model parameters using an analytical model that is similar to that used in \cite{vdBosch_04} and \cite{More_09a, More_09b, More_11}, and described below.
	
	\subsection{Galaxy number density}
	
	The total number density of galaxies in a given luminosity bin defined by $L_1$ and $L_2$ can be straightforwardly calculated from the CLF according to
	\begin{equation}\label{ngalpred}
		n_{\rm gal}(L_1, L_2) = \int\limits_{L_1}^{L_2} \int\limits_{M_{\rm min}}^\infty \Phi_{\rm tot}(L | M) \, n_\rmh (M) \, \mathrm{d}M \, \mathrm{d}L,
	\end{equation}
	where $n_\rmh$ is the halo mass function. In this work, we extract the mass function directly from the \texttt{SMDPL} $z = 0.05$ simulation. Specifically, any integral over the halo mass function is replaced by a sum over abundances of haloes in bins of $0.1 \ \mathrm{dex}$. As in the mocks, we assume a minimum halo mass of $M_{\rm min} = 3 \times 10^{10} \ h^{-1} M_\odot$ to host a galaxy.
	
	\subsection{Red fraction of primaries}
	
	We estimate the expectation value for the red fraction of primaries as the red fraction of centrals:
	\begin{equation}\label{fprirest}
		f_{\rm pri, r} (L_1, L_2) \approx 
		\frac{n_{\rm c, r} (L_1, L_2)}{n_{\rm c, r} (L_1, L_2) + n_{\rm c, b} (L_1, L_2)}\,.
	\end{equation}
	Here $n_{\rm c, r}(L_1, L_2)$ and  $n_{\rm c, b}(L_1, L_2)$  are the total number density of red and blue centrals, respectively, which we compute using equation~(\ref{ngalpred}), but with $\Phi(L|M)$ replaced with $f_\rmr(M) \Phi_{\rm c,r}(L|M)$ and $f_\rmb(M) \Phi_{\rm c,b}(L|M)$, respectively.
	
	This estimate underlies the assumptions that all primaries are centrals, and that the completeness of red primaries is identical to that of blue primaries. As we have shown in the middle panel of Fig.~\ref{fig:efficiency}, the impurity of our sample, defined as the fraction of primaries that are not centrals, is very low, typically below $1\%$. In addition, the left-hand panel of that figure shows that the completeness of centrals has only a small colour dependence. This is mainly due to fibre collisions which we correct for. In subsection \S\ref{subsec:accuracy_analytic} we show that equation~(\ref{fprirest}) does indeed give a good estimate for the red fraction of primaries.
	
	\subsection{Number of same-halo secondaries}
	\label{subsec:Nsamesec}
	
	Using the same arguments as for the red fraction of primaries, we can approximate the average number of same-halo secondaries around primaries by the number of satellites around centrals. Thus,
	\begin{align}
		&\langle N_\rms \rangle (L_1, L_2) \approx \nonumber\\ &\frac{\int\limits_{L_1}^{L_2}\int\limits_{M_{\rm min}}^\infty \langle N_\rms|M \rangle n_\rmh (M) \Phi_\rmc (L | M) f_{\rm ap} (L, M) \mathrm{d}M \mathrm{d}L}{n_\rmc (L_1, L_2)}
	\end{align}
	where $f_{\rm ap} (L, M)$ is the probability of halo members to fall within the aperture, i.e. inside the hollow cylinder defined by $R_\rms$, $R_\rmc$ and $(\Delta v)_\rms$, as discussed in \S\ref{subsec:cylinders}. Under the assumption that the primary is indeed the central, this probability is simply the expected fraction of satellites that lie in the cylinder centred on the halo. Given that $(\Delta v)_\rms$ is much bigger than the extent of the halo in redshift space, we have that
	\begin{equation}
		f_{\rm ap} (L, M) = 4 \pi \int\limits_0^{\rvir} \bar{n}_{\rm sat}(r|M)
		\left[\zeta(r, R_\rms(L)) - \zeta(r, R_\rmc)\right] \, r^2 \, \rmd r\,,
		\label{eq:f_ap}
	\end{equation}
	with $\bar{n}_{\rm sat}(r|M)$ the average radial profile of satellites around haloes of mass $M$, normalized such that
	\begin{equation}
		4 \pi \int\limits_0^{\rvir} \bar{n}_{\rm sat}(r|M) \, r^2 \, \rmd r = 1\,,
	\end{equation}
	and
	\begin{equation}
		\zeta(r, R) = \begin{cases}
			1 &\quad\text{if } r \leq R \\
			1 - \sqrt{1 - R^2 / r^2} &\quad\text{otherwise.} \\ 
		\end{cases}
	\end{equation}
	Note that this neglects the (small) possibility that some primaries are satellites. \cite{Lange_18a} have shown that this leads to negligible differences. Finally, $\bar{n}_{\rm sat}(r|M)$ is the average radial profile around halos of mass $M$, which we approximate as a generalized NFW profile (Equation~[\ref{nsatprof}]) with the median concentration, $c_{\rm vir}$, as measured in the appropriate halo catalogue.
	
	\subsection{Velocity dispersion}
	
	The observed velocity dispersion of same-halo secondaries can be calculated analogously to their number. We approximate
	\begin{align}
		\sigma^2 &(L_1, L_2) \approx \nonumber\\
		&\frac{\int\limits_{L_1}^{L_2}\int\limits_{M_{\rm min}}^\infty w (L, M) \, \sigma^2_{\rm ap} (L, M) \, n_\rmh (M) \, \Phi_\rmc (L | M) \, \mathrm{d}M \, \mathrm{d}L}{\int\limits_{L_1}^{L_2}\int\limits_{M_{\rm min}}^\infty w (L, M) \, n_\rmh (M) \, \Phi_\rmc (L | M) \, \mathrm{d}M \, \mathrm{d}L}.
	\end{align}
	The quantity $w(L,M)$ is a weight function that enforces either satellite-weighting 
	or host-weighting of the velocity dispersion. In the case of the satellite-weighted velocity  dispersion the weight function is
	\begin{equation}
		w_{\rm sw} (L, M) = f_{\rm ap} (L, M) \langle N_\rms|M \rangle,
	\end{equation}
	the expected number of satellites inside the aperture. For the host-weighted velocity dispersion the weight function, assuming a Poisson distribution for the satellite occupation, is
	\begin{equation}
		w_{\rm hw} (L, M) = 1 - \exp \left[ - f_{\rm ap} (L, M) \, \langle N_\rms|M \rangle \right],
	\end{equation}
	the probability of having at least one same-halo secondary inside the aperture. Finally, $\sigma^2_{\rm ap}$ is the average velocity dispersion of satellites inside the aperture and depends on $L$, and $M$. This can be calculated similarly to $f_{\rm ap} (L, M)$ in equation (\ref{eq:f_ap}) using equation~(\ref{eq:Jeans}),
	\begin{align}
		&\sigma^2_{\rm ap} (L, M) = \nonumber\\
		&\frac{\int\limits_0^{\rvir} \bar{n}_{\rm sat} (r | M) \, \sigma^2 (r | M) \, \left[\zeta(r, R_\rms(L)) - \zeta(r, R_\rmc)\right] \, r^2 \, \rmd r}{\int\limits_0^{\rvir} \bar{n}_{\rm sat} (r | M) \, \left[\zeta(r, R_\rms(L)) - \zeta(r, R_\rmc)\right] \, r^2 \, \rmd r}.
	\end{align}
	Again, we have neglected the spread in halo concentrations and have used the median halo concentration for each halo mass.
	
	\subsection{Accuracy of the analytic model}
	\label{subsec:accuracy_analytic}
	
	We can test the accuracy of our analytic model by comparing it to Monte-Carlo simulations. For the present test, we populate the SMDPL simulation with galaxies according to our default galaxy occupation model and a satellite phase-space distribution implied by $\gamma = \mathcal{R} = 1$; however, the results of this test are qualitatively very similar for the other two radial profiles. We create mock galaxy surveys according to the recipe described in \S\ref{sec:mocks}. We than measure all quantities using the procedure outlined in \S\ref{sec:observations} for making measurements on observational data. In this sense, our tests represent forward models of all observables. We repeat this exercise $1000$ times and calculate the average and dispersion for all observables among these different realizations. As discussed earlier, we only use those $\langle N_\rms \rangle$, $\sigma_{\rm hw}$ and $\sigma_{\rm hw}^2 / \sigma_{\rm sw}^2$ which could be measured for all $1000$ realizations, i.e. there were always at least an estimated $10$ satellites to calculate those quantities. Furthermore, we exclude measurements for $\sigma_{\rm hw}^2 / \sigma_{\rm sw}^2$ if $\langle N_\rms \rangle < 0.01$ on average. The reason is that the resulting ratios are very non-Gaussian and could bias our inferences if we were to approximate them as being Gaussian. In any way, $\sigma_{\rm hw} / \sigma_{\rm sw} \approx 1$ if $\langle N_\rms \rangle \ll 1$, so we do not expect these measurements to contain much information.
	\begin{figure*}
		\centering
		\includegraphics[width=\textwidth]{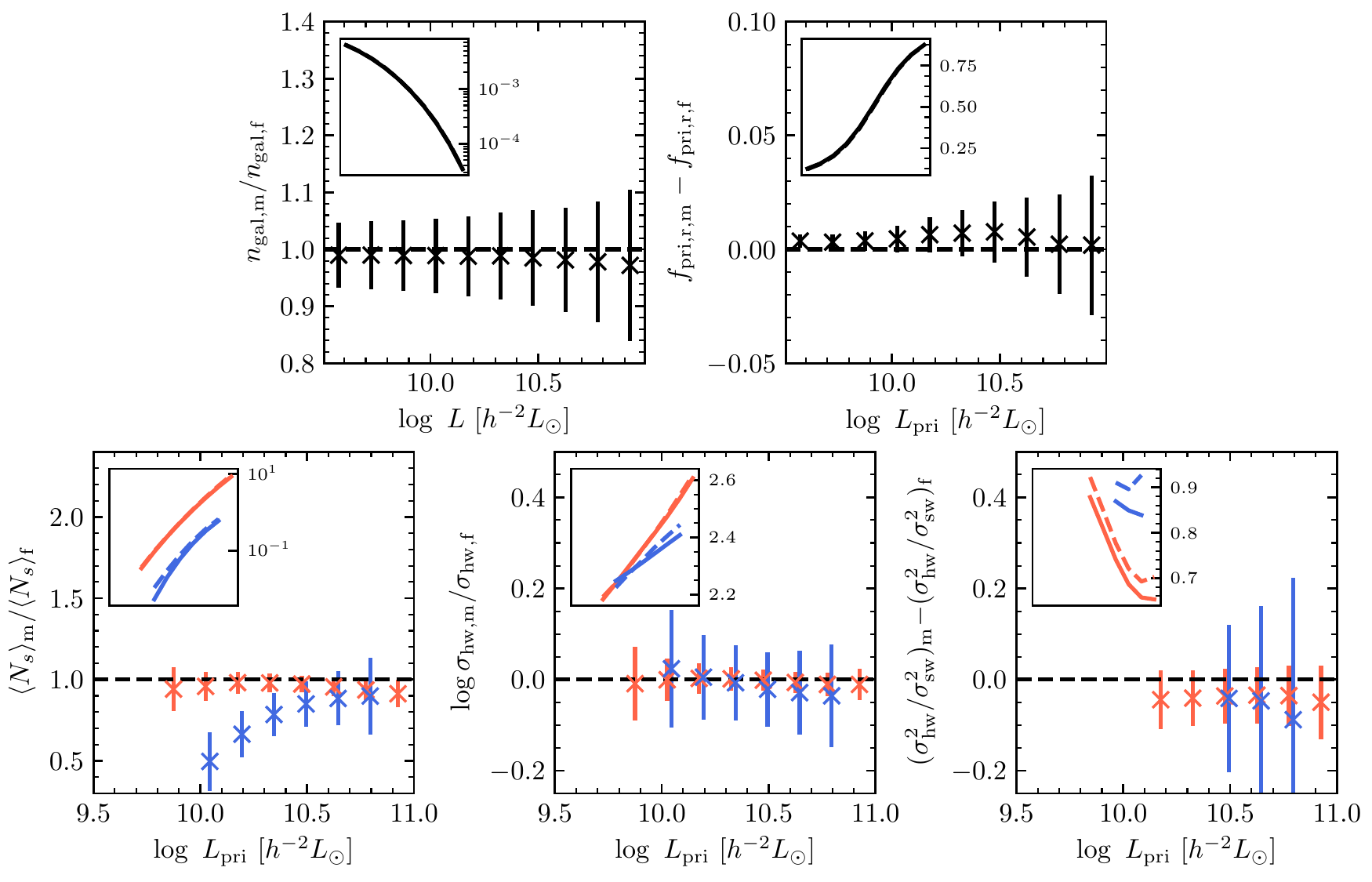}
		\caption{Comparison of the different observables derived from direct forward modelling and our simple analytical model. The main panels directly highlight differences between the two. Error bars show the $1 \sigma$ spread observed in $1000$ mock catalogues. The satellite observables in the lower row have been measured for red and blue primaries separately, as indicated by the colour of the error bars. In all cases, the dashed line indicates perfect agreement between the analytic model and the full forward-modelling approach. Finally, each smaller panel shows the absolute values of those observables for the analytic model (solid) and the forward-modelling result (dashed).}
		\label{fig:bias}
	\end{figure*}
	
	Figure \ref{fig:bias} compares the average values of the observables obtained in this forward-modelling fashion with the analytic predictions. The main panels directly show differences between the two, whereas the smaller panels show the absolute values. Error bars denote the $68\%$ scatter in those $1000$ mock catalogues. We first note that the overall qualitative agreement is good as all salient trends are recovered by the analytic prediction. In the following, we will briefly discuss some of the shortcomings.
	
	The number density of galaxies is basically recovered perfectly in the analytic model. This is not surprising as both the analytic model and the mock catalogues use the halo mass function of \texttt{SMDPL}. The analytic model generally overpredicts the red fraction of primaries. However, in each bin the difference is at most $\sim 1\sigma$. The reason is that red central galaxies reside, on average, in more massive haloes, which are more strongly clustered. This explains why red centrals have a slightly lower probably than blue centrals to be considered isolated (cf. left-hand panel of Fig.~\ref{fig:efficiency}). Also, the analytic model underpredicts the average number of same-halo secondaries for both red and blue primaries. The main reason for that is that our analysis procedure in \S\ref{sec:observations} does not perfectly reject interlopers. Particularly, we assume $P(\dv)$ of interlopers to be flat, whereas they are slightly more abundant at $\dv \sim 0$ \citep[see][]{vdBosch_04, More_09b}. Therefore, the membership probabilities for $\dv \sim 0$ secondaries and, accordingly, the overall number of satellites are overestimated. Given their much higher interloper fraction (cf. right-hand panel of Fig.~\ref{fig:efficiency}), it is also evident why this effect is strongest for faint blue primaries. Finally, the velocity dispersions are generally recovered very well, while the ratio of host- to satellite-weighted velocity dispersion is underestimated by at most around $\sim 0.5 \sigma$ in each bin.
	
	When using the analytic model to fit observations, we apply a correction to account for observational biases and biases in the model. This is discussed in more detail in the next section.
	
	\section{Analysis Procedure}
	\label{sec:analysis}
	
	When applying the analysis procedure described in \S\ref{sec:observations} to a spectroscopic catalogue we extract several observables. Particularly, we use $n_{\rm gal}$, $f_{\rm pri, r}$, $\langle N_\rms \rangle$, $\log \sigma_{\rm hw}$ and $\sigma_{\rm hw}^2 / \sigma_{\rm sw}^2$ as our data points from which we construct the data vector $\boldsymbol{D}$. Note that the latter $3$ quantities are measured for red and blue primaries separately. We have chosen $\log \sigma_{\rm hw}$ and $\sigma_{\rm hw}^2 / \sigma_{\rm sw}^2$, as opposed to for example $\sigma_{\rm hw}$ and $\sigma_{\rm sw}$, because their distributions in a large number of mocks with the same input models can be better, but not perfectly, described by Gaussian distributions.
	\begin{figure}
		\centering
		\includegraphics[width=\columnwidth]{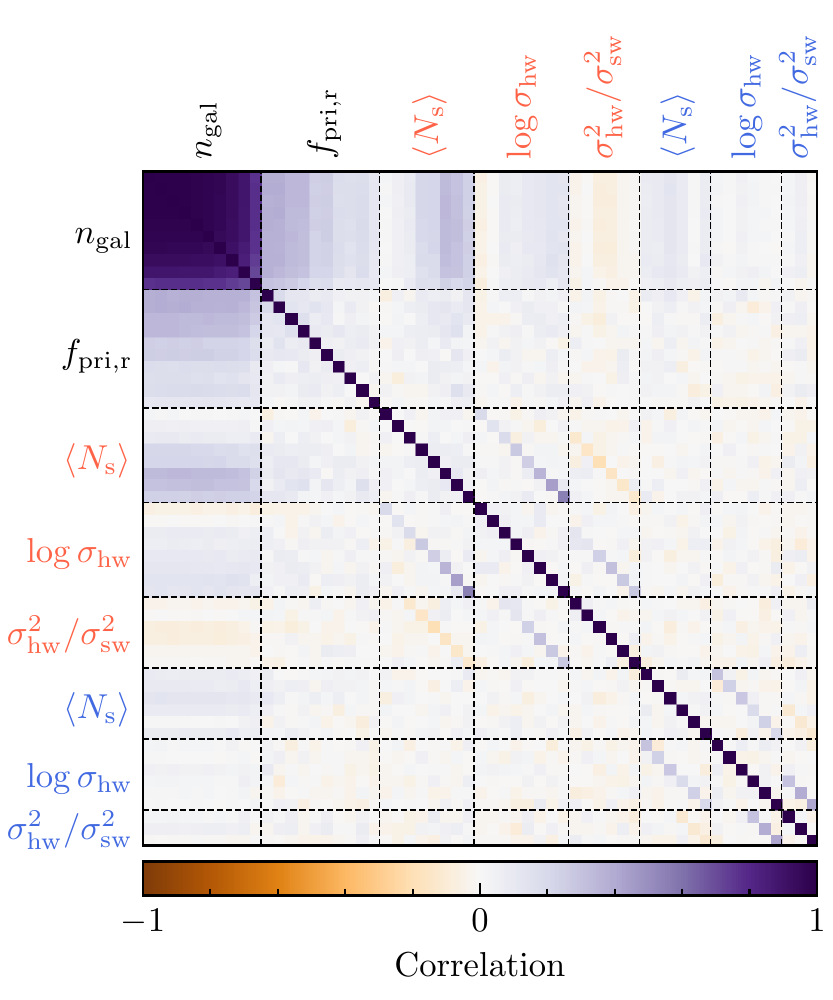}
		\caption{The correlation matrix of different observables when populating the mocks according to the default model and having satellites follow an NFW profile. The correlation matrix has been obtained from $1000$ mock catalogues. The dashed lines separate the different kinds of observables, whereas each single square represents this kind of observable for a certain bin in primary luminosity. The satellite statistics are measured for red and blue primaries separately, as indicated by the colour of the labels.}
		\label{fig:covariance}
	\end{figure}
	
	We construct covariance matrices using $1000$ mock catalogues derived from the same input model. In each mock the observer is placed at a different, random location within the simulation volume, thus accounting for sample variance. In this paper, the covariance matrix is always constructed using the default parameters described in Table \ref{tab:parameters}. An estimate $\boldsymbol{\hat{C}}$ for the covariance matrix is calculated via
	\begin{equation}
		\boldsymbol{\hat{C}} = \frac{1}{N_S - 1} \sum_i^{N_S} (\boldsymbol{D}_i - \langle \boldsymbol{D} \rangle) (\boldsymbol{D}_i - \langle \boldsymbol{D} \rangle)^t,
	\end{equation}
	where $N_S = 1000$ is the number of samples, $\boldsymbol{D}_i$ is the data vector of the $i$-th mock and $\langle \boldsymbol{D} \rangle$ the average for all mock catalogues. In Figure \ref{fig:covariance} we show the correlation matrix for all observables. For this particular matrix, we used mocks were satellites follow an NFW profile, $\gamma = \mathcal{R} = 1$. Covariance matrices for other profiles are very similar. Note that we only show observables that can be measured for all mocks.
	
	The first thing to note is the very strong correlation for the $n_{\rm gal}$ measurements in different bins. This well known cosmic variance \citep[see e.g.][]{Moster_11, Smith_12} implies that the overall number density of galaxies of all luminosities has a large uncertainty, whereas the ``shape'' of the luminosity function is very well constrained. In fact, when fitting a model to observational data or mock catalogues, we will add an additional $2\%$ error, i.e. $(0.02 n_{\rm gal})^2$, to the diagonal elements of the luminosity function covariance matrix. This reduces the strong covariance and allows the abundances of galaxies of different luminosities to vary independently. We do this primarily to not bias our inference on the galaxy-halo connection in case our parametrization, as described in \S\ref{sec:galaxy-halo_connection}, cannot perfectly reproduce the observed luminosity function. Next, the first column in Figure \ref{fig:covariance}, which depicts the correlation of $n_{\rm gal}$ with other observables, shows that $n_{\rm gal}$ is weakly, positively correlated with $\langle N_\rms \rangle$ and $\log \sigma_{\rm hw}$, and weakly anti-correlated with  $\sigma_{\rm hw}^2 / \sigma_{\rm sw}^2$. These trends are expected because higher values of $n_{\rm gal}$ should be associated with over-dense regions. Consequently, the average dark matter halo masses will be slightly higher \citep[e.g.,][]{Mo_04}, even at fixed primary luminosity. Thus, $\langle N_\rms \rangle$ and $\log \sigma_{\rm hw}$ are expected to increase and $\sigma_{\rm hw}^2 / \sigma_{\rm sw}^2$ to decrease. Finally, there is also a correlation of $\langle N_\rms \rangle$ with $\log \sigma_{\rm hw}$ and an anti-correlation of $n_{\rm gal}$ and $\sigma_{\rm hw}^2 / \sigma_{\rm sw}^2$ at fixed colour and luminosity of the primary. If all primaries of a given colour and luminosity live in slightly more massive haloes, we expect $\langle N_\rms \rangle$ and $\log \sigma_{\rm hw}$ to be higher and $\sigma_{\rm hw}^2 / \sigma_{\rm sw}^2$ to be lower.
	
	We note that $\sigma_{\rm hw}$ and $\sigma_{\rm sw}$ in the same primary bin are always highly positively correlated (not shown). For example, for low-luminosity samples with primaries having only $0$ or $1$ secondaries, $\sigma_{\rm hw} \equiv \sigma_{\rm sw}$. Thus, there should be no extra information in measuring $\sigma_{\rm sw}$ over just $\sigma_{\rm hw}$ alone. In their analysis, however, \cite{More_09b, More_11} implicitly assumed those measurements to be independent. Thus, especially for low-luminosity primaries where $\sigma_{\rm hw}$ and $\sigma_{\rm sw}$ are highly correlated, this should lead to an underestimation of the uncertainty in the average halo mass. On the other hand, the uncertainties in the spread in halo mass at fixed luminosity which is characterized by the ratio of $\sigma_{\rm hw}$ and $\sigma_{\rm sw}$ is likely overestimated. Thus, we expect better constraints on the scatter in luminosity at fixed halo mass, $\sigma_{\rm r}$ and $\sigma_{\rm b}$ by using the full covariance matrix.
	
	The precision matrix $\boldsymbol{\Psi}$ is the inverse of the covariance matrix $\boldsymbol{C}$. In case of a noisy estimate of the latter, the unbiased estimator $\boldsymbol{\hat{\Psi}}$ for the precision matrix becomes
	\begin{equation}
		\boldsymbol{\hat{\Psi}} = \frac{N_S - N_D - 2}{N_S - 1} \, \boldsymbol{\hat{C}}^{-1},\label{eq:precision}
	\end{equation}
	where $N_D$ is the number of data points \citep{Taylor_13}.
	
	We have shown in \S\ref{sec:model} that our analytic model has some small biases with respect to the full-forward modelling approach for a given set of parameters. We assume that those biases do not change substantially across all models allowed by the data. Under this assumption, we attempt to correct for those systematic offsets by shifting the model predictions as
	\begin{equation}
		\boldsymbol{M}^\star (\boldsymbol{\theta}) = \boldsymbol{M}(\boldsymbol{\theta}) + \boldsymbol{B}(\boldsymbol{\theta_0}),
	\end{equation}
	where
	\begin{equation}
		\boldsymbol{B}(\boldsymbol{\theta_0}) = \langle \boldsymbol{D} \rangle(\boldsymbol{\theta_0}) - \boldsymbol{M}(\boldsymbol{\theta_0})
	\end{equation}
	is the bias between the average data vector of the $1000$ mock catalogues, $\langle \boldsymbol{D} \rangle(\boldsymbol{\theta_0})$ and the (uncorrected) analytic model,
	$\boldsymbol{M}(\boldsymbol{\theta_0})$, 
	for the default parameter set, $\boldsymbol{\theta_0}$.
	
	Given a set of observables $\boldsymbol{D}$, we calculate the likelihood via
	\begin{equation}
		\mathcal{L} \propto \exp \left[ - \frac{(\boldsymbol{M}^\star (\boldsymbol{\theta}) - \boldsymbol{D})^t \boldsymbol{\Psi} (\boldsymbol{M}^\star (\boldsymbol{\theta}) - \boldsymbol{D})}{2} \right] = \exp\left[ -\chi^2 / 2\right].
		\label{eq:likelihood}
	\end{equation}
	We use \texttt{MultiNest} \citep{Feroz_08, Feroz_09, Feroz_13} to calculate the resulting posterior. We assume flat priors in all parameters with the ranges shown in Table \ref{tab:parameters}. We run \texttt{MultiNest} with $10,000$ live points and a target sampling efficiency of $0.5\%$. Constant efficiency mode is turned off. We use $\Delta \ln \mathcal{Z} = 10^{-4}$ as a stopping criterion, where $\Delta \ln \mathcal{Z}$ is the uncertainty in the estimate of the global Bayesian evidence. We confirmed that our results are converged by running \texttt{MultiNest} with $20,000$ live points and a target efficiency of $0.2\%$. This exercise yielded very similar posteriors.
	
	Finally, we note that we have created an estimate for the covariance matrix and the bias for the default parameter set in Table \ref{tab:parameters}. This makes sense since the mock catalogues that we test our method on, as described in the next section, have the same parameters. However, when applying our method to observations, the input parameters are unknown. This problem can be solved in an iterative fashion. One can start out with the default model used here, calculate an estimate for the bias and the covariance and use this to get a new best-fit model. Afterwards, one then calculates a new bias and covariance for this new model and so on. We find that this algorithm converges sufficiently after only $3$ iterations when applied to SDSS (Lange et al., in preparation). In appendix \ref{sec:iteration}, we test the iteration scheme on mock catalogues. We note that such an iteration is only necessary if the initial model is not a good fit to the data. As long as a good fitting model is chosen, the estimate for the bias and covariance are rather stable.
	
	\section{Application to Mocks}
	\label{sec:application_mocks}
	
	We now validate our analysis procedure by applying it to mock catalogues with known input parameters. In all cases, those input values are the ones listed in Table~\ref{tab:parameters}.
	
	\subsection{Mocks with known satellite phase-space distributions}
	\label{subsec:mocks_simple}
	
	First, we apply our analysis procedure to mock catalogues where satellites obey analytical radial density profiles and the spherically averaged Jeans equation without anisotropy, as described in \S\ref{subsec:phase-space_distributions}. Furthermore, we assume that we know the true underlying radial profile of satellites perfectly. Thus, we always use the same radial profile to model the data as we used to create the mock catalogue in question.
	\begin{figure*}
		\centering
		\includegraphics[width=\textwidth]{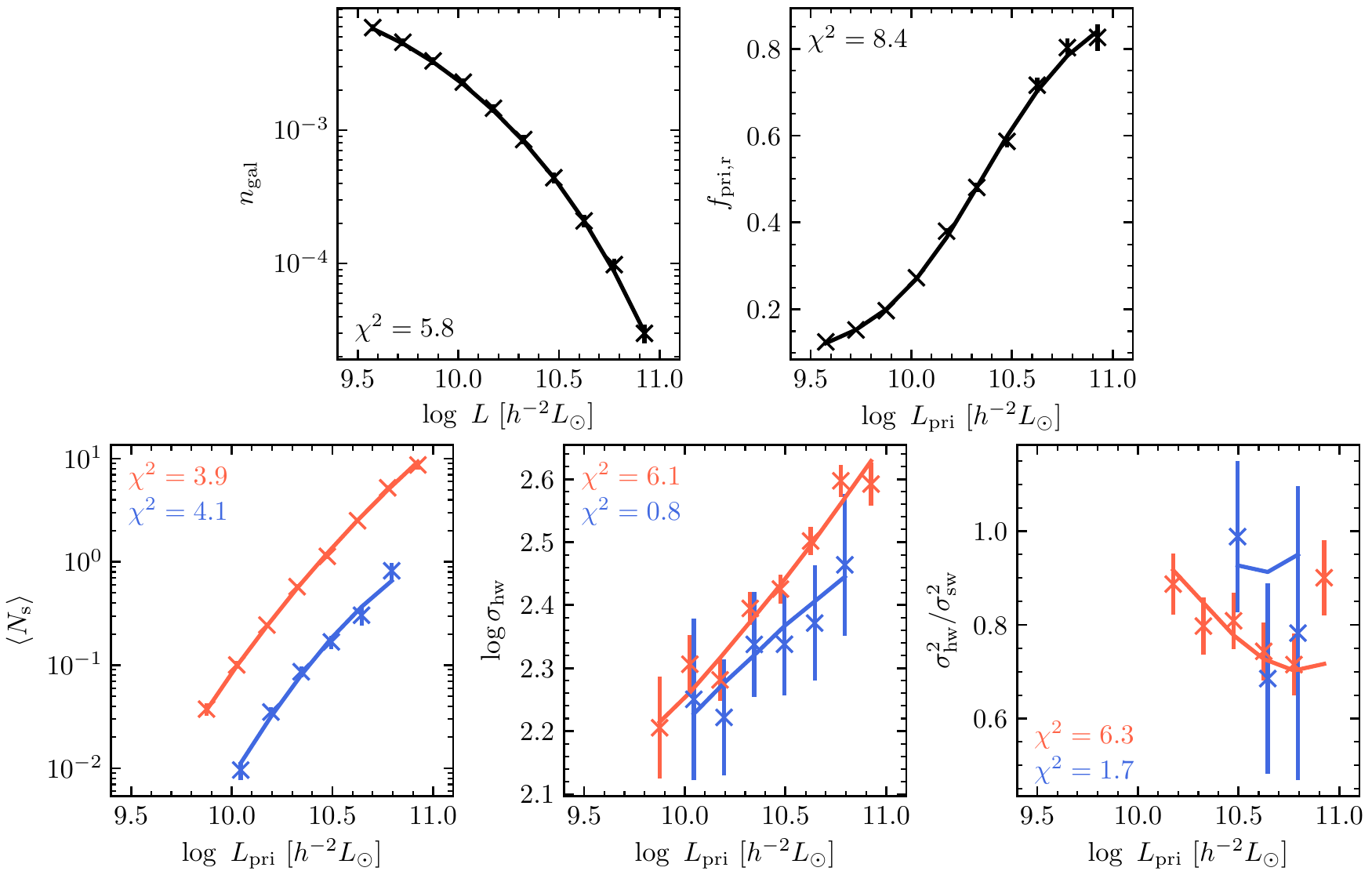}
		\caption{The prediction of the best-fitting model against the measured observables in a mock catalogue where satellites follow an NFW profile and predictions of the isotropic Jeans equation. Error bars denote the $1\sigma$ uncertainty derived from mock catalogues. In each panel, we also show the $\chi^2$ of each set of observables. Note however that there is a weak correlation between different observables and the total $\chi^2$ is not the sum of the individual $\chi^2$ values.}
		\label{fig:nfw_constraints_analytic}
	\end{figure*}
	
	\begin{figure}
		\centering
		\includegraphics[width=\columnwidth]{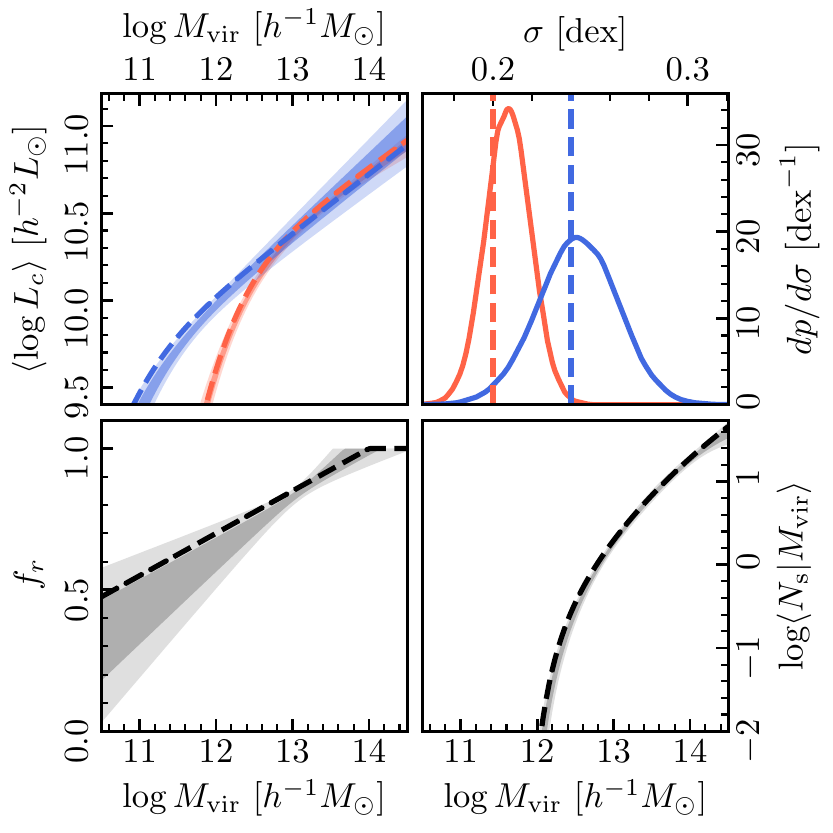}
		\caption{The posterior predictions for the galaxy-halo connection when analysing a mock catalogue where satellite galaxies follow an NFW profile and obey the isotropic Jeans equation. Bands show the $1$ and $2\sigma$ regions and the dashed lines input values. The solid lines in the upper right panel is a Gaussian kernel density estimate of the posterior. We show the average luminosity as a function of halo mass for red and blue centrals (top, left), the scatter in luminosity at fixed halo mass (top, right), the red fraction of centrals as a function of mass (bottom, left) and the satellite occupation (bottom, right). Note that we eliminated satellites that are brighter than centrals. Thus, the actual average satellite occupation will have a small dependence on the colour of the central. We have omitted that complication for clarity and calculated the satellite occupation as if satellites brighter than centrals were not removed.}
		\label{fig:model_post_nfw}
	\end{figure}
	
	In Figure \ref{fig:nfw_constraints_analytic} we show the best-fit model when fitting to a mock SDSS catalogue where satellites follow an NFW profile, i.e. $\mathcal{R} = \gamma = 1$. As shown, we can fit all observations in the mock catalogue very well with an overall $\chi^2$ of $35$ for $54 - 16 = 38$ degrees of freedom. The posterior predictions for the galaxy-halo connection inferred from the same mock catalogue are shown in Figure \ref{fig:model_post_nfw}. We show the input model as dashed lines and error bands denote the $68\%$ and $95\%$ posterior prediction. We see that this analysis succeeded in recovering the input model with all posteriors including the input to within $\sim 2 \sigma$. We have also looked at all $16$ free model parameters and their $16 \times 15 / 2 = 120$ two-dimensional distributions and also found a good agreement. Finally, we have applied our analysis procedure to mocks produced with the other two radial profiles, $\gamma = 1$, $\mathcal{R} = 2.0$ and $\gamma = 0$, $\mathcal{R} = 2.5$, and were also able to recover the input parameters to within reasonable uncertainties.
	
	\subsection{Mocks with unknown phase-space distributions}
	
	So far, we have assumed that the underlying radial profile of satellites is known a priori. This will not be the case for actual observations as there is still debate whether satellite galaxies follow the dark matter distribution \citep[e.g.,][]{vdMarel_00, vdBosch_05c, Tal_12, Cacciato_13, Guo_15a} or are more radially biased \citep[e.g.,][]{Yang_05, Chen_08, More_09b, Hoshino_15}.
	\begin{figure*}
		\centering
		\includegraphics[width=\textwidth]{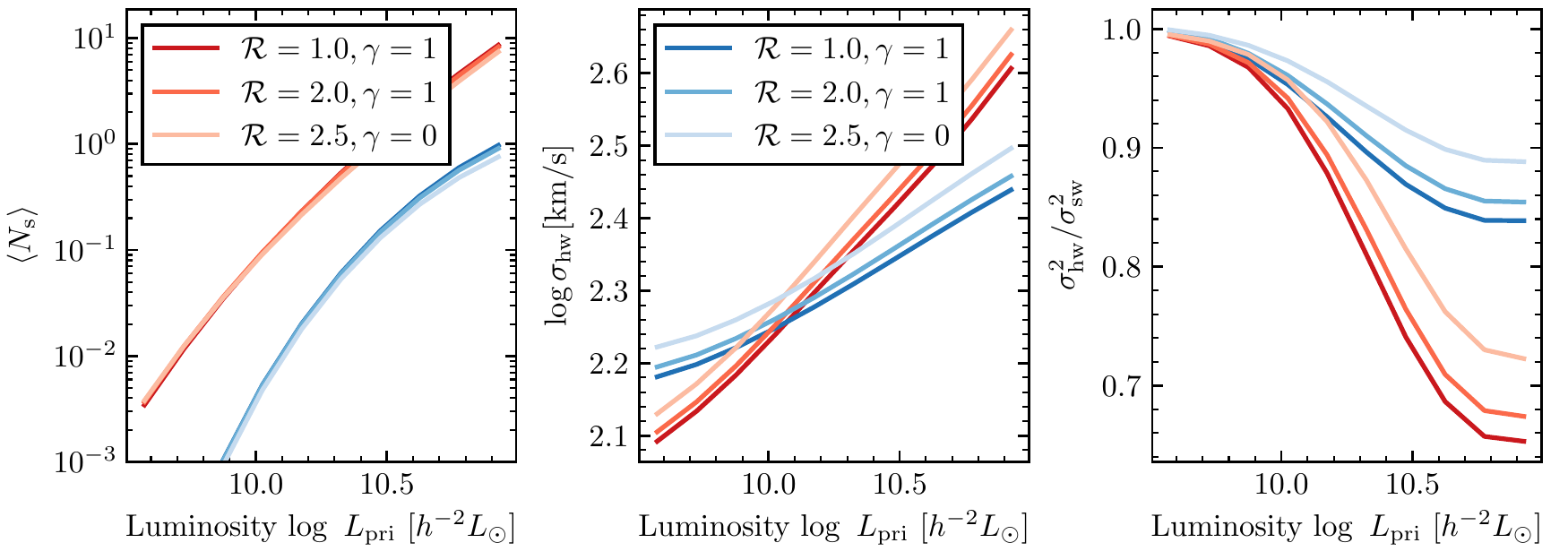}
		\caption{The dependence of the constraints used in this work on the radial profile if satellites obey the spherically averaged Jeans equation without anisotropy. All observables have been predicted from the analytic model. In each panel, we show the predictions with lines of different saturation. Darker lines indicate more radially anti-biased profiles. The number density of galaxies and the red fraction of primaries are not shown since they have no direct dependence on the radial profile of satellites.}
		\label{fig:profile_vs_observables}
	\end{figure*}
	
	We show in Figure \ref{fig:profile_vs_observables} how the different observables are impacted by the choice of the radial profile. For this plot, all parameters describing the galaxy occupation are fixed to their default values. Only $\mathcal{R}$ and $\gamma$ describing the radial profile of satellites have been changed. The observables have been predicted from the analytical model. The results from the forward-modelling approach are very similar but cannot be inferred for all observables at all primary luminosities. The number density of galaxies and the red fraction of primaries are not shown in this comparison as they have no dependence on the radial profile of satellites. Note that in the forward-modelling approach, the red fraction of primaries does have a dependence on the radial profile due to misidentification of satellites as primaries. But that effect is insignificant.
	
	Several of our observables are sensitive to the choice of satellite distribution. The number of satellites inside the aperture generally decreases for more extended radial profiles, e.g. $\mathcal{R} = 2.5$ and $\gamma = 0$. The only exception is for low-luminosity primaries where a substantial amount of satellites are within $60 \kpch$ and not counted due to potential issues with fibre collisions, as discussed in \S\ref{subsec:fibre_collisions}. The velocity dispersion always increases for more extended radial profiles, as expected from Figure \ref{fig:profile_vs_dispersion}. The difference is $\sim 0.05 \ \mathrm{dex}$ between an unbiased profile, $\mathcal{R} = \gamma = 1$, and the most radially anti-biased profile, $\mathcal{R} = 2.5$ and $\gamma = 0$, irrespective of the luminosity or colour of the primary. Finally, the ratio of host- and satellite-weighted velocity dispersion is generally lower for more radially concentrated profiles. The reason is that for those profiles the high-mass halos \textit{at a fixed luminosity} have a larger fraction of their satellites inside the aperture, thereby increasing the satellite-weighted velocity dispersion. Altogether, the different radial profiles predict significantly different observables for a fixed galaxy-occupation model. Thus, choosing a realistic model for the spatial distribution of satellites is important in order to draw unbiased inferences regarding the galaxy-halo connection.
	\begin{figure*}
		\centering
		\includegraphics[width=\textwidth]{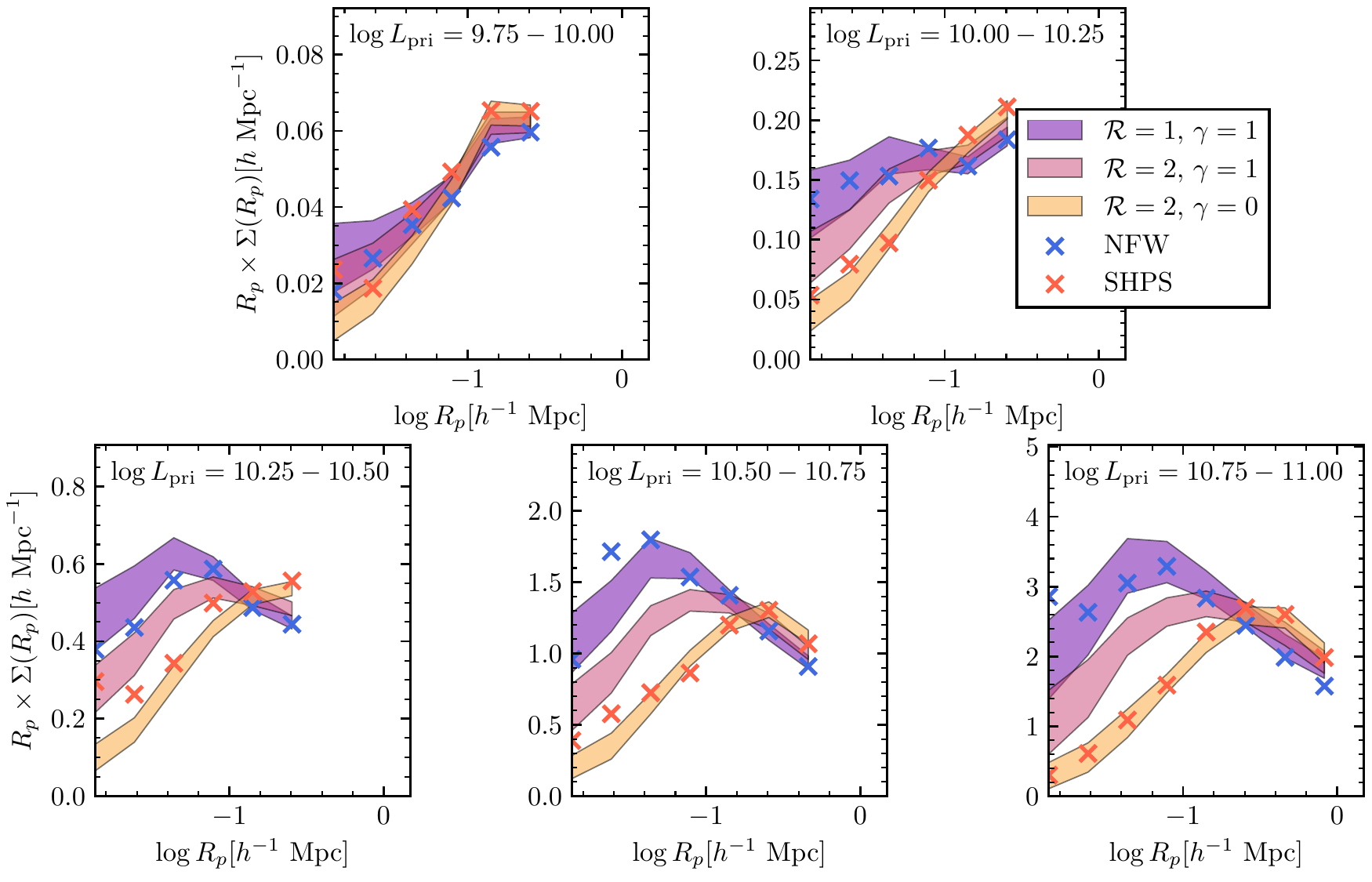}
		\caption{The observed radial profile in different mock catalogues. In each panel we show the projected surface density of secondaries around primaries in a given luminosity bin, as indicated in the top right corner. We only include secondaries with $|\Delta z| < 3 \sigma$, where $\sigma$ is given by equation (\ref{eq:sigma_cylinder}). The shaded bands show the $68\%$ spread of $100$ mock catalogues with different input radial profiles for satellites. The crosses indicate the values obtained for the specific mock catalogue that is being analysed We show a mock catalogue where satellite galaxies follow the best-fit NFW profile of the dark matter (blue, NFW) and a catalogue where those galaxies follow the phase-space distribution of subhaloes (red, SHPS).}
		\label{fig:profile_observed}
	\end{figure*}
	
	So far, we have not used observations that would directly constrain the radial profile of satellites. Additionally, we keep $\gamma$ and $\mathcal{R}$ fixed when fitting the data.  However, upon obtaining model fits for all three different radial profiles using our analysis techniques, we can create mock catalogues and directly compare the resulting observed radial profiles to the observations.
	
	Let us assume the radial profile of the mock catalogue analysed in the previous subsection was unknown. In Figure \ref{fig:profile_observed}, we show the projected surface number density of secondaries around primaries of different luminosities, where the blue data points show the data from this specific mock catalogue. We compare this to mock catalogues with known radial distributions of satellites and the same parameters for the galaxy--halo connection as bands. Those bands denote the $68\%$ range in the mocks due to random fluctuations. We make no attempt to correct for interlopers, instead we only consider secondaries with $|\Delta z| < 3 \sigma$, where $\sigma$ is given by equation (\ref{eq:sigma_cylinder}). A correction for fibre collisions has been applied in all mock catalogues and secondaries with $R_\rmp < 60 \kpch$ are also included in this analysis.
	
	Based on this analysis of the projected surface densities of secondaries, we would infer that galaxies in this mock catalogue follow the NFW profile of the dark matter, i.e. $\mathcal{R} = \gamma = 1$. Thus, we would only consider the best-fitting parameters and uncertainties for the galaxy-halo connection under this assumption. Thus, as in the previous subsection, we would get a good agreement between the derived and the input parameters for the galaxy--halo relation. Note that when analysing SDSS, the input parameters are unknown and we would instead compare projected surface densities to those from mocks derived with the best-fitting parameters. However, this has no significant impact on the surface number densities.
	
	What would happen if we would assume the wrong radial profile when analysing the mock catalogue? To test this we fitted the above mock catalogue with $\mathcal{R} = \gamma = 1$ with a model where $\mathcal{R} = 2$ and $\gamma = 1$. We find that the resulting shift in the posterior prediction for the galaxy--halo connection is very modest. For example, the one-dimensional posteriors for $\log L_0$, $\log M_1$, $\gamma_2$ and the luminosity scatter $\sigma$ for both red and blue centrals all change by less than $0.2 \sigma$.
	
	Finally, we note that when analysing the radial profile we assumed that BHGs always reside exactly at their host halo centres. This assumption is not true in general and one might worry that violations of this assumption alter the observed radial profile of secondaries around primaries, i.e. BHGs. However, \cite{Lange_18a} have shown that the effect on the projected number density of secondaries is negligible for the purposes of this analysis.
	
	\subsection{Mocks with unknown and complex phase-space distributions}
	\label{subsec:mocks_complex}
	
	We have so far assumed that satellites have a spherically isotropic distribution and obey the Jeans equation without anisotropies. In many ways, this is a gross simplification. For example, subhaloes and satellite populations can have various degrees of non-sphericity and there is a substantial amount of substructure in phase-space. Furthermore, dark matter haloes are not fully relaxed. Thus, there is no a priori reason to assume that they would obey the Jeans equation \citep{Ye_17, Wang_18b}. Additionally, we have neglected higher order moments of the velocity distribution and velocity anisotropy in our calculation. Here, we test whether or not these complications influence our inferences regarding the galaxy-halo connection at a level that is relevant compared to statistical uncertainties.
	
	To evaluate this we create a mock catalogue where satellite galaxies are placed on resolved dark matter subhaloes in SMDPL. This addresses all the potential issues discussed above because the subhalo distributions within individual host halos are not spherically symmetric and their velocities distributions exhibit deviations from the form assumed in a Jeans analysis. We use the \texttt{SubhaloPhaseSpace} module of \texttt{halotools} to place satellites on subhaloes. We first determine for each dark matter halo the number of satellites that it hosts, according to the recipe described in \S\ref{sec:galaxy-halo_connection} and regardless of how many dark matter subhaloes we actually find in SMDPL. Then we place the satellites on those subhaloes with the highest $M_{\rm peak}$, the maximum dark matter halo mass achieved over the lifetime of each subhalo. In rare cases where we have more satellites than subhaloes in the same halo, we use the relative phase-space positions of random subhaloes hosted by haloes of a similar mass. We then proceed to generate a mock SDSS-like catalogue and analyse it in the same way as the one in the previous section.
	
	We start by analysing the radial profile of satellites, as shown by the red crosses in Figure \ref{fig:profile_observed}. When comparing the mock catalogue to other mock catalogues with analytical phase-space profiles, we see that it most closely resembles the most radially anti-biased profile, $\mathcal{R} = 2.5, \gamma = 0$. This is another manifestation of the well-known result that subhaloes are spatially anti-biased with respect to dark matter \citep[see e.g.][]{Diemand_04, vdBosch_16}. Thus, for our posterior prediction on the galaxy--halo connection, we would choose results when assuming $\mathcal{R} = 2.5, \gamma = 0$. In this case, we find a good fit with $\chi^2 = 36$ for $57 - 16 = 41$ degrees of freedom.
	
	\begin{figure}
		\centering
		\includegraphics[width=\columnwidth]{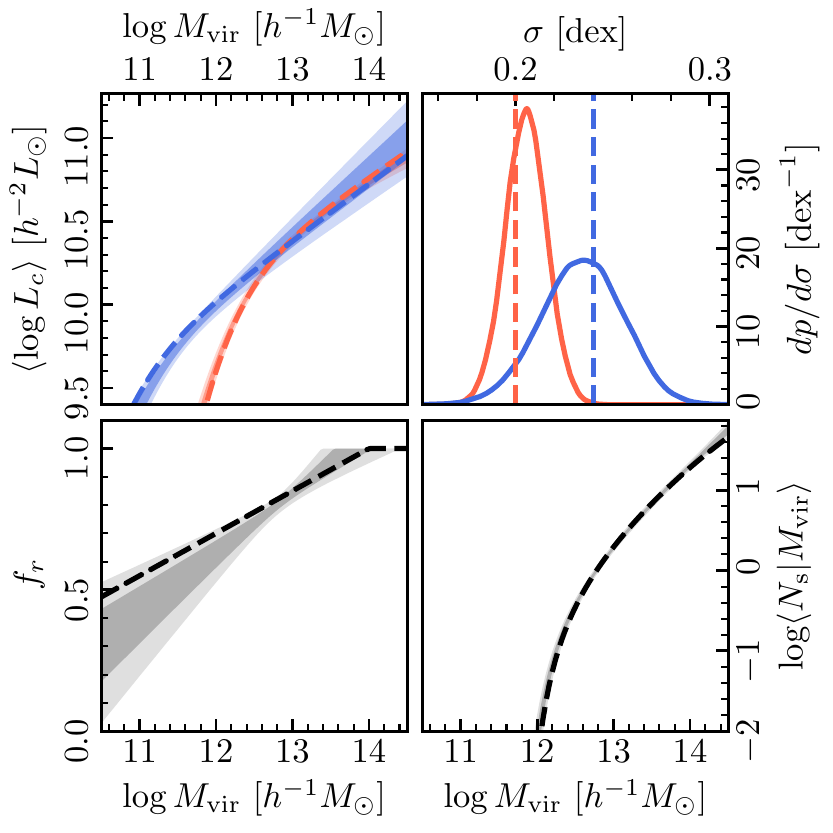}
		\caption{Similar to Figure \ref{fig:model_post_nfw}, but now showing the posterior predictions for the galaxy-halo connection when analysing a mock catalogue where satellite galaxies are place on subhaloes. Bands show the $1$ and $2\sigma$ regions and the dashed lines input values. The solid lines in the upper right panel is a Gaussian kernel density estimate of the posterior.}
		\label{fig:model_post_shps}
	\end{figure}
	Using this mock catalogue, we can evaluate the impact of orbital anisotropy on our velocity dispersion measurements. We find that independent of luminosity satellites have an anisotropy parameter of around $\beta \sim 0.3$, indicating slightly radially biased orbits. As discussed in \cite{vdBosch_04}, the velocity dispersion of all satellites inside $r_{\rm vir}$ is expected to be almost independent of $\beta$. However, because our cylindrical isolation criterion samples satellites within small projected separations $R_p$, the line-of-sight velocity is prefentially aligned with the radial velocity component. The calculations in \cite{vdBosch_04} suggest that this preference, coupled with $\beta \sim 0.3$, could lead to a percent level increase in $\sigma_{\rm los}^2$. Indeed, in the same mocks we find that, irrespective of primary luminosity, $\sigma_{\rm los}^2$ is at most $\sim 2\%$ larger than $\sigma_{\rm 3D}^2 / 3$, where $\sigma_{\rm 3D}^2$ is the three-dimensional velocity dispersion. This shows that orbital anisotropy has a negligible effect on our measurements and can safely be negelected in the modelling.
	
	In Figure \ref{fig:model_post_shps} we show the posterior predictions when analysing the above mentioned mock catalogue assuming $\mathcal{R} = 2.5, \gamma = 0$. This Figure is analogous to Figure \ref{fig:model_post_nfw}. Also, similar to Figure \ref{fig:model_post_nfw}, we find a good agreement of our posterior prediction with the input model. We have repeated this above experiment two more times (not shown), finding similar results.
	\begin{figure}
		\centering
		\includegraphics[width=\columnwidth]{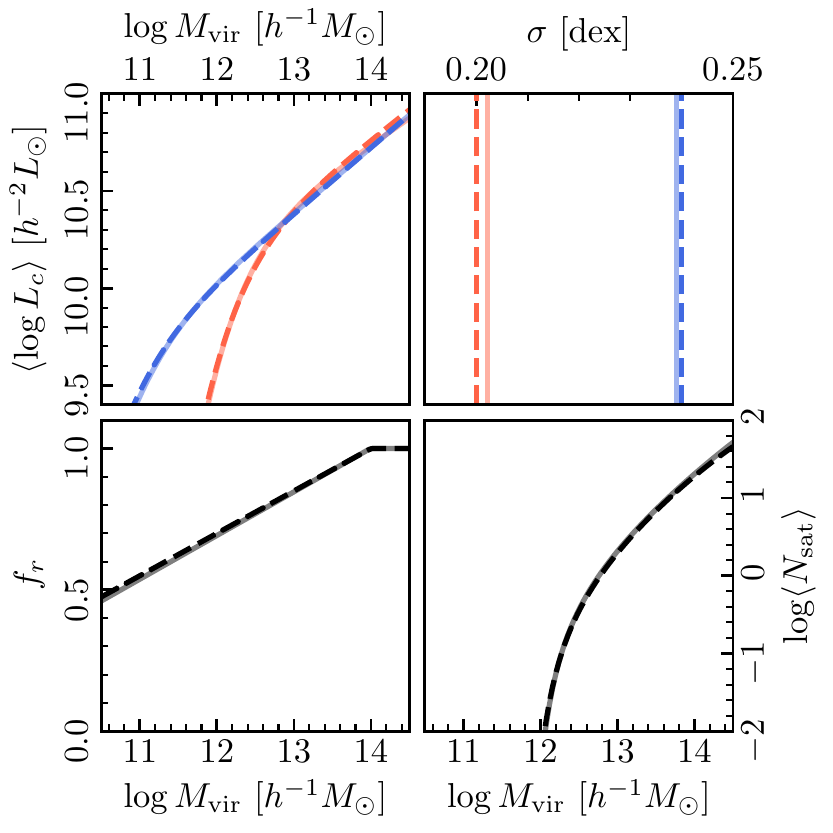}
		\caption{Similar to Figure \ref{fig:model_post_shps}, but this time we show the best-fit model (solid lines) when analysing 100 mock catalogues where satellites are placed on subhaloes.}
		\label{fig:default_vs_fit_100}
	\end{figure}
	
	So far, we have shown a reasonable agreement between posterior predictions and input values for the galaxy--halo connection when analysing single SDSS-like mock catalogues, e.g. Figure \ref{fig:model_post_shps}. However, it is still possible that our predictions are biased by $1$ or $2\sigma$. Due to statistical fluctuations, such a bias would not necessarily show up when only performing a single analysis. We address this issue by fitting our model to $100$ mock catalogues simultaneously. As usual, each mock catalogue uses a different random realization of the default CLF model, and a random position and orientation for the observer. For each mock catalogue, we extract the data vector from which we can calculate the $\chi^2$ with respect to a model according to Eq.~(\ref{eq:likelihood}). To calculate the total likelihood of each model in the posterior estimation we add the $\chi^2$ of all $100$ mock catalogues, i.e. $\mathcal{L} \propto \exp [-\sum \chi^2 / 2]$. The result is shown in Figure \ref{fig:default_vs_fit_100}. Since we simulate a fit to $100$ statistically independent mock catalogues, the posterior uncertainties are mostly negligible. Thus, we only show the best-fit model. We find that it is in excellent agreement with the input model. Note that the disagreement for the scatter in luminosity for red and blue centrals is still within the statistical uncertainties when analysing $100$ mock catalogues. Altogether, we find that all biases in our analysis when analysing a single mock catalogue should be within the statistical uncertainty.
	
	\subsection{Competitiveness of Constraints}
	
	Let us further investigate how competitive our constraints are. \cite{Guo_15b} constrain the halo occupation distribution (HOD) of galaxies above $M_{\rm r}^{0.1} < -19$, corresponding to roughly $L_\rmr^{0.1} > 10^{9.5} h^{-1} M_\odot$, from redshift-space clustering. Among other things, they constrain the satellite fraction in this sample to within $\pm 1.0\%$. Similarly, \cite{Sinha_18} using projected clustering and group catalogues of the same sample obtained constraints of around $\pm \sim 1.5\%$. On the other hand, in our mock analysis we achieve constraints of the order of $\pm 0.8\%$. Other quantities are difficult to compare with due to the different functional forms of the HOD parametrizations used by these studies and the CLF employed here. \cite{Cacciato_13} constrained both the galaxy-halo connection and cosmological parameters using observations of the luminosity function, galaxy clustering and galaxy-galaxy lensing in the range $0.011 \leq z \leq 0.245$ and $\log L_{\rm r}^{0.1} > 8.7$ in SDSS. For example, they were able to constrain the scatter in luminosity at fixed halo mass to within $\pm 0.007 \ \mathrm{dex}$. This is comparable to our constraint on the scatter for red centrals of $\pm 0.012 \ \mathrm{dex}$. Finally, \cite{Cacciato_13} constrain the median luminosity at $\log M / (\Msunh) = 13$ to within $\sim 0.02 \ \mathrm{dex}$, similar to our constraints. Note however that this comparison is only approximate given that \cite{Cacciato_13} also constrained cosmological parameters and that our mock catalogues might be substantially different from the actual Universe. Nevertheless, overall our constraints using only satellite kinematics and galaxies with $0.02 \leq z \leq 0.067$ and $\log L > 9.5$ seem to be competitive to results from a combined analysis of clustering and lensing, redshift space clustering or group catalogues.
	
	\subsection{Additional Caveats}
	
	Altogether, we have shown that certain simplifications in our model, particularly assuming the radially symmetric Jeans equation without anisotropy, seem to have a very small effect on our inferred galaxy--halo relation.
	
	However, other simplifications to the model may also induce small biases in inferred galaxy--halo connection parameters, particularly assumptions regarding our primary galaxy samples. One example is the assumption that all BHGs are centrals. This assumption may cause small, but insignificant biases in our inferences \citep{Lange_18a}. Further, the reader should be aware that we only analyse a subset of all centrals (compare left panel of Fig. \ref{fig:efficiency}). This occurs because our isolation criterion will likely exclude centrals in high density regions. Therefore, if the systems excluded by this criterion exhibit systematically different kinematics or a systematically different galaxy--halo relation, our inferred parameters will be biased relative to the global galaxy--halo relation. In previous work, \cite{Faltenbacher_10} have shown that the dynamics of subhaloes vary systematically with environment at fixed halo mass. Haloes at fixed mass in low density regions have a lower subhalo velocity dispersion. Confirming the results of \cite{Faltenbacher_10}, we find that haloes whose centrals are identified as primaries have a slightly lower velocity dispersion for their subhaloes. However, this effect is only significant for lower halo masses and practically disappears for $M_{\rm vir} > 10^{13} \ h^{-1} M_\odot$. Additionally, this effect is already included in the mock catalogues where satellites are placed on subhaloes. Finally, given how competitive our constraints are compared to a combination of clustering and lensing \citep{Cacciato_13}, it seems likely that our results are also impacted by cosmological parameters \citep[see also][]{Li_12}. However, a framework that also allows for variations in the cosmology is beyond the scope of this work.
	
	\section{Conclusion}
	\label{sec:conclusion}
	
	Satellite kinematics are a potentially powerful probe of the galaxy--halo connection, which have hitherto been rarely utilized, except for massive clusters with large numbers of satellites. We developed a new method to analyse satellite kinematics that aims to rectify this shortcoming. This method represents a continuation of the work by \cite{vdBosch_04} and \cite{More_09b, More_09a, More_11}. We test and validate our analysis framework using mock catalogues of increasing complexity. We demonstrate the need to accurately account for interlopers and fibre-collision-induced incompleteness in the spectroscopic survey. For the time being, practical limitations on computational resources necessitate the use of a semi-analytical method to model satellite kinematics; however, wider use of forward models for satellite kinematics is a high priority for future work. In this work, we use forward-modelling to construct covariance matrices, and to validate and calibrate our analytical methods. We highlight a few improvements with respect to \cite{More_11}, which, as we demonstrate in Paper~II, alleviate the tension between their results and alternative constraints on the galaxy--halo connection from galaxy abundances, galaxy-galaxy lensing, clustering, and group catalogues.
	
	\begin{itemize}
		
		\item We have shown in \S\ref{subsec:fibre_collisions} that fibre collisions bias the measured velocity dispersion low by $\sim 5\%$ if not corrected. Furthermore, fibre collisions also lead to biased inferences of the radial profiles of satellites, which further alter the expected relation between velocity dispersion and halo mass. Both effects will likely lead to an underestimation of the average halo mass at fixed central luminosity. We have introduced a framework to correct for fibre collisions in \S\ref{subsec:fibre_collisions}. We have demonstrated that this method is highly effective in correcting the spectroscopic incompleteness for all observables considered in this work.
		
		\item \S\ref{sec:analysis} describes a framework to specifically correct for the biases introduced by the analysis pipeline and the analytic model. Specifically, we use detailed mock SDSS-like mock catalogues including the effects of fibre collisions and the survey mask to calibrate our model. For example, contrary to the results by \cite{More_09b}, we do not find that the velocity dispersion of satellites can be extracted in an unbiased manner in the presence of interlopers. Instead, we find the ``measured'' velocity dispersion to be biased low, similar to the results by \cite{Becker_07}.
		
		\item By using detailed mock catalogues we are also able to create realistic covariance matrices that include the non-negligible correlations between different observables. For example, the host-weighted and satellite weighted velocity dispersion estimates $\sigma_{\rm hw}$ and $\sigma_{\rm sw}$ are highly correlated. Thus, their ratio is much better constrained than if they were statistically independent. Since this ratio is a measure of scatter in halo mass at fixed luminosity \citep{More_09a}, this should lead to much stronger constraints on the latter compared to the results by \cite{More_09b, More_11}.
		
		\item We now use the galaxy luminosity function as a constraint. As shown by \cite{Li_12}, the combination of number density and average velocity dispersion can constrain, for example, the scatter in galaxy luminosity or stellar mass at fixed halo mass.
		
	\end{itemize}
	
	We have tested our method by applying it to detailed mock catalogues with increasing complexity. One might worry that our assumption that satellites obey the spherically symmetric Jeans equation might bias our inferences. For example, subhaloes that host satellites are known to have radial anisotropy, phase-space substructure and are not spherically symmetric. However, we have demonstrated that this does not bias our inferences, at least in the case of $M_{\rm peak}$ selected subhaloes.
	
	We have also demonstrated that the constraints derived from our analysis are competitive with respect to studies utilizing galaxy clustering, galaxy-galaxy lensing or a combination thereof. A more detailed analysis will be conducted in Paper II by applying the framework developed here to SDSS. Beyond constraining a traditional mass-dependent CLF model, there are many exciting applications of satellite kinematics that would also benefit from an improved understanding of satellite kinematics. For example, galaxy clustering might be affected by assembly bias \citep{Zentner_14, Zentner_16}, the dependence of clustering on halo properties other than mass. While it is not clear whether or not satellite kinematics will be significantly influenced by this, naive theoretical considerations suggest that they may. Thus, the combined analysis of satellite kinematics and clustering might further strengthen our observational constraints on the galaxy--halo relationship, particularly assembly bias. As an even more ambitious goal, satellite kinematics may be used to constrain cosmological parameters in a manner that is quite distinct from traditional probes \citep{Li_12}. The quantity and quality of data are rapidly improving, so the tools that we use to interpret data must mature at a commensurate pace if we are to make the most of this data. This work is the first step toward using the kinematics of galaxies on nonlinear scales to interpret galaxy surveys and use them to inform the galaxy--halo connection and constraint cosmology.
	
	\section*{Acknowledgements}
	
	The work presented in this paper has greatly benefited from discussions with Surhud More and Andrew Hearin.
	
	FvdB and JUL are supported by the US National Science Foundation (NSF) through grant AST 1516962. ARZ and KW are funded by the Pittsburgh Particle Physics, Astrophysics, and Cosmology Center (Pitt PACC) at the University of Pittsburgh and by the NSF through grant AST 1517563. ASV is funded by Pitt PACC and the NSF through grant AST 1516266. This research was supported  by the HPC facilities operated by, and the staff of, the Yale Center for Research Computing, and in part by the NSF under grant PHY 1125915 and PHY 1748958. FvdB received additional support from the Klaus Tschira foundation, and from the National Aeronautics and Space Administration through Grant No. 17-ATP17-0028 issued as part of the Astrophysics Theory Program.
	
	This work made use of the following software packages: \texttt{matplotlib} \citep{Hunter_07}, \texttt{SciPy}, \texttt{NumPy} \citep{vdWalt_11}, \texttt{Astropy} \citep{Astropy_13}, \texttt{Cython} \citep{Behnel_11}, \texttt{halotools} \citep{Hearin_17a}, \texttt{Corner} \citep{Foreman-Mackey_16}, \texttt{MultiNest} \citep{Feroz_08,Feroz_09, Feroz_13}, \texttt{PyMultiNest} \citep{Buchner_14}, \texttt{mangle} \citep{Hamilton_04, Swanson_08} and \texttt{pymangle}\footnote{\url{https://github.com/esheldon/pymangle}}. We also thank the open-source developers behind \texttt{Ubuntu}, \texttt{GNOME}, \texttt{Xfce}, \texttt{Spyder}, \texttt{JabRef}, \texttt{TexStudio} and \texttt{Terminator}. All the above mentioned software packages helped to greatly expedite this work.
	
	The authors gratefully acknowledge the Gauss Centre for Supercomputing e.V. (www.gauss-centre.eu) and the Partnership for Advanced Supercomputing in Europe (PRACE, www.prace-ri.eu) for funding the MultiDark simulation project by providing computing time on the GCS Supercomputer SuperMUC at Leibniz Supercomputing Centre (LRZ, www.lrz.de).The Bolshoi simulations have been performed within the Bolshoi project of the University of California High-Performance AstroComputing Center (UC-HiPACC) and were run at the NASA Ames Research Center.
	
	\bibliographystyle{mnras}
	\bibliography{bibliography}
	
	\appendix
	\section{Effects of the SDSS Tiling Algorithm}
	\label{sec:tiling}
	
	As detailed in \cite{Blanton_03a}, the SDSS fibre spectrograph can at most observe $592$ science targets on a single spectroscopic plate. Due to the overall large-scale structure of galaxies and the resulting large variance of target density, a naive uniform placement of these tiles on the sky would lead to a significant spectroscopic incompleteness in high density regions. To circumvent this problem, the SDSS tiling algorithm perturbs the tile positions slightly such that the fraction of decollided targets that receive spectroscopic redshifts is maximised. Particularly, this results in high density regions on the sky having a higher density of tiles and a larger fraction of overlapping tile areas. Although this optimal set of tiles is constructed using only the decollided targets \citep{Blanton_03a}, it results in a higher spectroscopic completeness for potentially collided galaxies, i.e. galaxies not in the decollided set, in overdense regions.

	\begin{figure}
		\centering
		\includegraphics[width=\columnwidth]{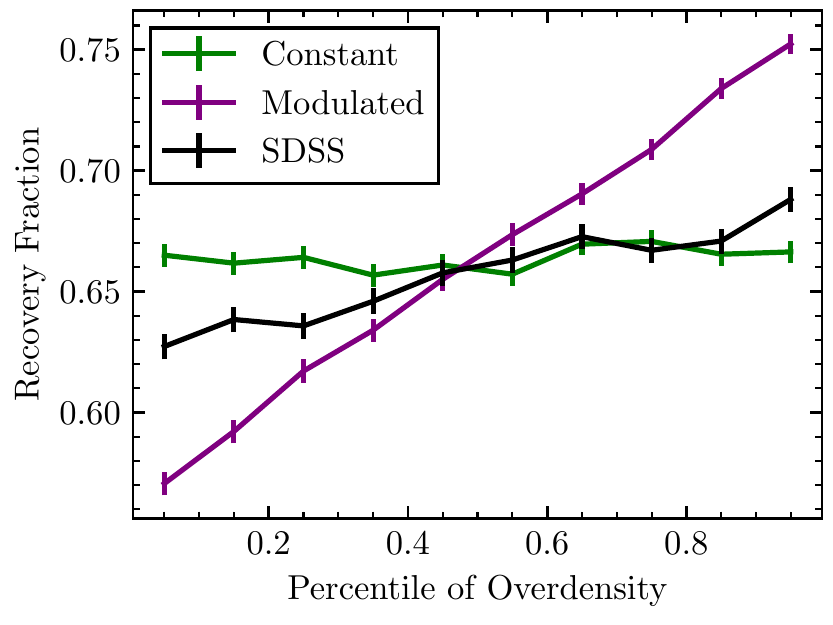}
		\caption{The recovery fraction, defined as the fraction of targets with other targets within $55''$ that receive spectroscopic redshifts, as a function of density. The density is defined as the number of targets within $1.5 \ \mathrm{deg}$. We show the results of SDSS (black) against those of a mock catalogue with our standard implementation of fibre collisions (green) and an alternative scheme (purple).}
		\label{fig:recovery_fraction}
	\end{figure}
	
	Implementing the entire SDSS tiling algorithm in our mock catalogues is beyond the scope of this work. We can however estimate its impact by comparing with the actual SDSS data. In Figure \ref{fig:recovery_fraction} we show the fraction of targets with neighbours within $55''$ that receive spectroscopic redshifts. We plot this recovery fraction as a function of the overall large scale density. Particularly, we calculate for each object the number of other targets within $1.5 \ \mathrm{deg}$, i.e. the size of the SDSS spectroscopic plates. The recovery fraction is plotted as a function of the overall percentile of this overdensity with respect to all other potentially collided targets. For the SDSS, we see a roughly linear dependence with the recovery fraction increasing from $\sim 63\%$ to $\sim 69\%$ when going from underdense to overdense regions. We tested other radii for measuring the density but found similar or smaller dependences.
	
	This overall scaling of the recovery fraction with large-scale density is not included in our default mock catalogues. The concern is that average halo masses will be positively correlated with this overall large-scale density. Particularly, our fibre collision correction method implicitly assumes the recovered galaxies to be representative of those lost due to fibre collisions. However, the results in Figure \ref{fig:recovery_fraction} suggest that the SDSS tiling algorithm will be slightly biased towards resolving collisions (i.e., assigning fibres to collided galaxies) for high-mass haloes, which, on average, reside in denser environments. We can address this concern using mock catalogues. For our regular mocks we assumed that $35\%$ of all potentially collided targets receive spectroscopic redshifts, irrespective of the overall large-scale density. The resulting recovery fraction is shown by the green line in Figure \ref{fig:recovery_fraction}. In a different set of mocks, we assume this fraction to linearly increase with the rank of the overdensity on $1.5 \ \mathrm{deg}$ from $15\%$ to $55\%$. The resulting recovery fraction is shown by the purple line and clearly more pronounced than the SDSS data. We create $100$ mock catalogues where for each mock we run the two different fibre collision algorithms. Afterwards, we analyse these mocks with our analysis pipeline and compare the differences in the observables, e.g. the velocity dispersion, between these two sets of mocks. We do not find any significant difference in the observables for these two kind of mock making algorithms. This suggests that the overall modulation of the fibre collision recovery fraction with large-scale density is negligible and does not impact our results.
	
	\section{Iterative Bias and Covariance Estimation}
	\label{sec:iteration}
	
	Throughout \S\ref{sec:application_mocks} we have used the bias and covariance estimate derived from the input model. However, when analysing any galaxy survey, this input model is unknown. Instead, one can estimate the bias and covariance from the best-fit model. Since the parameters of this model also depend on the bias and covariance, an iterative scheme has 	to be used.
	
	\begin{figure}
		\centering
		\includegraphics{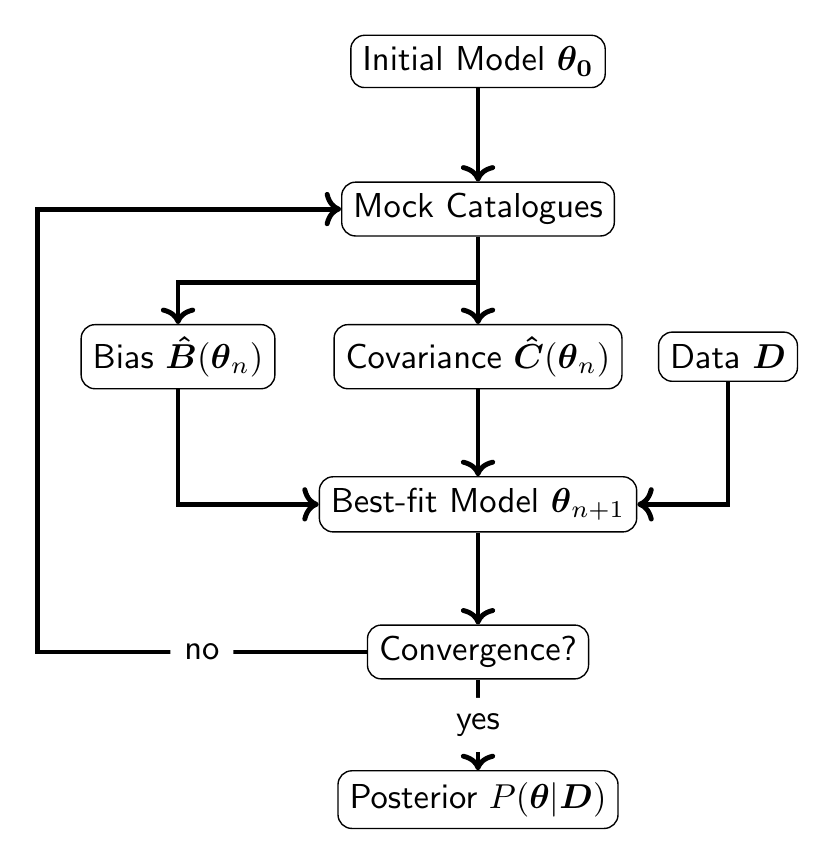}
		\caption{Diagram outlining our analysis procedure of the SDSS data. We start out with an initial model $\boldsymbol{\theta_0}$ and create mock catalogues. These mock catalogues are used to estimate the covariance of the data and the offset between the analytical model for the constraints and the forward-modelling results. Using these estimates, the analytical model and the observations, a new best-fitting model $\boldsymbol{\theta_1}$ is obtained. This process is repeated until the sequence of best-fitting models $\boldsymbol{\theta}_{0, 1, ..., n}$ is reasonably converged.}
		\label{fig:graph}
	\end{figure}
	
	Figure \ref{fig:graph} outlines the general procedure. We start out with an initial guess for the galaxy--halo connection given by $\boldsymbol{\theta_0}$. We then create mock catalogues for this particular choice of parameters and construct the bias vector $\boldsymbol{\hat{B}}$, precision and covariance matrix from them. These are then used to find a new best-fit model $\boldsymbol{\theta}_1$ using the analytic model plus the bias correction that maximizes the likelihood. From this model we create another series mock catalogues and re-compute the bias and covariance. This process is repeated until a convergence criterion is reached, signalling that the best-fit models do not change significantly. 
	
	How do we asses the convergence of this algorithm? We have a series of best-fit models $\boldsymbol{\theta}_{0,1,...,n}$ based upon different estimates for the bias and covariance. We now take the latest estimate for the latter and compute the $\chi^2$ of all $\boldsymbol{\theta}_{0,1,...,n}$. By construction, $\boldsymbol{\theta}_n$ will have the lowest $\chi^2$. On the other hand, previous iterations will 	have higher values because they used different estimates for the bias and covariance for the minimization. If the posterior of $\boldsymbol{\theta}$ were described by a multivariate Gaussian distribution with $16$ degrees of freedom, $68\%$ of all models would lie within $\Delta \chi^2 = 18$ of the best-fit model. Thus, if $\chi^2 \left(\boldsymbol{\theta}_{n - 1} \right) - \chi^2 \left( \boldsymbol{\theta}_n \right) < 18$, the best-fitting model did not change significantly from the previous iteration and we regard the result as converged.
	
	Here, we test the iteration scheme using the mock catalogue analysed in \S\ref{subsec:mocks_simple}. Instead of choosing an arbitrary starting model $\boldsymbol{\theta}_0$, we choose a starting bias and covariance. For the bias, we simply assume no bias, i.e. $\boldsymbol{B}(\boldsymbol{\theta}_0) = 0$. Furthermore, we assume a diagonal covariance matrix with a $5\%$ error on $n_{\rm gal}$ and $\langle N_\rms \rangle$, a $1\%$ error on $f_{\rm pri, r}$ and an error of $0.05$ for $\log \sigma_{\rm hw}$ and $(\sigma_{\rm hw}^2 / \sigma_{\rm sw}^2)$. Convergence is achieved at the third iteration. The resulting posterior predictions are virtually indistinguishable from the ones presented in \S\ref{subsec:mocks_simple}. For example, none of the $16$ one-dimensional posteriors shifts by more than $0.3 \sigma$ in the mean.
	
	\label{lastpage}
\end{document}